\documentclass[twocolumn]{aastex631}
\shorttitle{Retrieval of Gl 229B}
\shortauthors{Calamari et al.}

\graphicspath{{./}{figures/}}

\usepackage{amsmath}
\usepackage{mathrsfs}
\usepackage{natbib}
\usepackage{xcolor}

\begin{document}

\title{An Atmospheric Retrieval of the Brown Dwarf Gliese 229B}

\author[0000-0002-2682-0790]{Emily Calamari}
\affiliation{The Graduate Center, City University of New York,
New York, NY 10016, USA}
\affiliation{Department of Astrophysics, American Museum of Natural History, New York, NY 10024, USA}

\author[0000-0001-6251-0573]{Jacqueline K. Faherty}
\affiliation{Department of Astrophysics, American Museum of Natural History, New York, NY 10024, USA}

\author[0000-0003-4600-5627]{Ben Burningham}
\affiliation{Centre for Astrophysics Research, School of Physics, Astronomy and Mathematics, University of Hertfordshire, Hatfield AL10 9AB}

\author[0000-0003-4636-6676]{Eileen Gonzales}
\affiliation{Department of Astronomy and Carl Sagan Institute, Cornell University, 122 Sciences Drive, Ithaca, NY 14853, USA}
\altaffiliation{51 Pegasi b Postdoctoral Fellow}

\author[0000-0001-8170-7072]{Daniella Bardalez-Gagliuffi}
\affiliation{Department of Physics \& Astronomy, Amherst College, 25 East Drive, Amherst, MA 01003, USA}
\affiliation{Department of Astrophysics, American Museum of Natural History, New York, NY 10024, USA}

\author[0000-0003-0489-1528]{Johanna M. Vos}
\affiliation{Department of Astrophysics, American Museum of Natural History, New York, NY 10024, USA}

\author[0000-0002-8871-773X]{Marina Gemma}
\affiliation{Department of Earth and Environmental Sciences, Columbia University, New York, NY 10027, USA}
\affiliation{Department of Earth and Planetary Sciences, American Museum of Natural History, New York, NY 10024, USA}

\author[0000-0001-8818-1544]{Niall Whiteford}
\affiliation{Institute for Astronomy, University of Edinburgh, Blackford Hill, Edinburgh, EH9 3HJ, UK}

\author{Josefine Gaarn}
\affiliation{Centre for Astrophysics Research, School of Physics, Astronomy and Mathematics, University of Hertfordshire, Hatfield AL10 9AB}

\begin{abstract}
We present results from an atmospheric retrieval analysis of Gl 229B using the \textit{Brewster} retrieval code. We find the best fit model to be cloud-free, consistent with the T dwarf retrieval work of \citet{Line2017}, \citet{Zalesky2022} and \citet{Gonzales2020}. Fundamental parameters (mass, radius, log($L_{\rm Bol}$ /$L_{\rm Sun}$), log(g)) determined from our model agree within 1$\sigma$ to SED-derived values except for $T_{\rm eff}$  where our retrieved $T_{\rm eff}$  is approximately 100 K cooler than the evolutionary model-based SED value. We find a retrieved mass of $50^{+12}_{-9}$ $M_{\rm Jup}$ , however, we also find that the observables of Gl 229B can be explained by a cloud-free model with a prior on mass at the dynamical value, 70 $M_{\rm Jup}$ . We are able to constrain abundances for H$_2$O, CO, CH$_4$, NH$_3$, Na and K and find a supersolar C/O ratio as compared to its primary, Gl 229A. We report an overall subsolar metallicity due to atmospheric oxygen depletion but find a solar [$C/H$], which matches that of the primary. We find that this work contributes to a growing trend in retrieval-based studies, particularly for brown dwarfs, toward supersolar C/O ratios and discuss the implications of this result on formation mechanisms, internal physical processes as well as model biases.
\end{abstract}

\section{Introduction} \label{sec:intro}
With masses $\leq$ 75 $M_{\rm Jup}$ , brown dwarfs are a category of astronomical objects whose core temperatures are too low to maintain stable hydrogen fusion throughout their lifetimes \citep{Chabrier1997}. Unlike main sequence stars, brown dwarfs will contract and cool as they age and progress through their spectral classification sequence \citep[M, L, T, Y;][]{Burgasser06, Kirkpatrick2005, Cushing2011}. These objects are often seen as a bridge between stars and planets as they are luminous enough to be directly imaged yet cool enough to have molecular-rich, and even condensate-rich, atmospheres, similar to what we see in Jupiter and large, gaseous exoplanets \citep[e.g.][]{Helling2014, Marley1997, Marley2002, Marley2013}. One path forward to understanding the breadth of exoplanet atmospheres, and the physical processes within, is by proxy through brown dwarfs.

While important work has been done to characterize isolated brown dwarfs through computational approaches \citep[e.g.][]{Line2017, Zalesky2022, Burningham2021}, in this work, we add to the list of retrieved companion objects by focusing on Gl 229B, the first discovered methane-bearing brown dwarf \citep{Oppenheimer1995, Nakajima1995} and a widely separated companion to a main sequence M-dwarf star, Gl 229A. Brown dwarfs that exist in co-moving pairs or systems, particularly ones with main-sequence stars, have become critical benchmarks in order to further establish our understanding of substellar mass objects. Discoveries of these types of systems are essential to brown dwarf science since we can use information from the primary star to place constraints on fundamental parameters for the system as a whole \citep[e.g.][]{Faherty2010, Faherty2021, Faherty2020, Pinfield2012, Kirkpatrick2001, Burningham2009, Burningham2011, Burningham2013, Dupuy2009}. Such works have used chemical abundances, activity and/or kinematics of the primary star to place metallicity, mass and age constraints on the companion. However, inherent to these constraints is the assumption that these companion objects formed together via the same formation mechanism which can bias our models and results.

\begin{deluxetable}{lcr}
\tablenum{1}
\tablecaption{Properties of Gliese 229B}
\tablewidth{0pt}
\tablehead{\colhead{Parameter} & \colhead{Value} & \colhead{Reference}}
\startdata
\hline
Spectral Type & T7p & 2\\
\hline
\multicolumn{3}{c}{\textbf{Astrometry}} \\
\hline
R.A. & $06^h10^m34.61^s$ & 1\\
Dec & $-21^h51^m52.66^s$ & 1\\
$\pi$ (mas) & 173.57 $\pm$ 0.017 & 1\\
$\mu_{\alpha}$ (mas yr$^{-1}$) & -135.692 $\pm$ 0.011 & 1\\
$\mu_{\delta}$ (mas yr$^{-1}$) & -719.178 $\pm$ 0.017 & 1\\
\hline
\multicolumn{3}{c}{\textbf{Photometry}} \\
\hline
$Y_{MKO}$ (mag) & 15.17 $\pm$ 0.1 & 3\\
$J_{MKO}$ (mag) & 14.01 $\pm$ 0.05 & 4\\
$H_{MKO}$ (mag) & 14.36 $\pm$ 0.05 & 4\\
$K_{MKO}$ (mag) & 14.36 $\pm$ 0.05 & 4\\
$L'_{MKO}$ (mag) & 12.24 $\pm$ 0.05 & 5\\
$M'_{MKO}$ (mag) & 11.74 $\pm$ 0.11 & 5\\
\hline
\multicolumn{3}{c}{\textbf{SED-Derived Parameters}} \\
\hline
Radius ($R_{\rm Jup}$ ) & $0.94 \pm 0.15$ & 6\\
Mass ($M_{\rm Jup}$ ) & $41 \pm 24$ & 6\\
log(g) & 4.96 $\pm$ 0.46 & 6\\
$T_{\rm eff}$  (K) & $927 \pm 79$ & 6\\
log($L_{\rm Bol}$ /$L_{\rm Sun}$) & $-5.21 \pm 0.05$ & 6\\
\enddata
\tablerefs{(1) \citet{Gaia2021}, (2) \citet{Burgasser06}, (3) \citet{Hewett06}, (4) \citet{Leggett2002b}, (5) \citet{Golimowski2004}, (6) \citet{Filippazzo2015}}
\label{tab:SED results}
\end{deluxetable}

Brown dwarf atmospheric research was initially grounded in the use of grid models (radiative-convective equilibrium atmosphere models) whose predicted spectra are fit to observed spectra in order to derive fundamental properties of a particular object such as temperature, gravity and metallicity \citep[e.g.][]{Burrows1993, Burgasser2007}. However, due to the complexity of these grid models as well as the complexity and diversity of brown dwarf atmospheres, this method has been shown to produce discrepancies in parameter estimation and fitted synthetic spectra \citep[e.g.][]{Cushing2008, Rice2010, Manjavacas2014}. In an attempt to constrain previously estimated fundamental parameters of Gl 229B, we turn to atmospheric retrievals, a spectral inversion technique that compliments the work of forward grid models by using minimal physical assumptions to determine more precise parameter values than are inferred by comparing synthetic to observed spectra.

For this retrieval work, we use \textit{Brewster}, a flexible framework composed of a forward model and a retrieval model. The forward model reproduces the object spectrum based on a combination of pre-determined and retrieved parameters while the retrieval model tests the goodness of fit of those parameters determined from the forward modelling and effectively proposes new parameter values \citep{Burningham2017}. This technique allows us to place constraints on gas abundances, thermal profile and, in the case of cloud models, cloud location and opacity.

In Section \ref{sec:literature}, we review published values of the Gl 229 system. In Section \ref{sec:SED}, we discuss the data used in constructing a spectral energy distribution (SED) as well as subsequent retrieval models and review SED-derived fundamental parameters for Gl 229B. In Section \ref{sec:brewster}, we discuss the retrieval framework and settings used in our work. In Section \ref{sec:retrieval}, we present retrieval results for Gl 229B. In Section \ref{sec:discussion}, we compare fundamental parameters derived from our retrieval model to literature predictions and SED-derived parameters. In Section \ref{sec:t comp}, we place our work in the context of previous retrieval work on T dwarfs. Finally, in Section \ref{sec:chemistry} we discuss retrieved C/O ratio and metallicity of Gl 229B and compare its retrieved chemistry to that of its primary, Gl 229A.

\begin{deluxetable*}{clcrr}[ht!]
\tablenum{2}
\tablecaption{Data Used in Retrieval Models and SED}
\label{tab:retrieval data}
\tablewidth{0pt}
\tablehead{
\colhead{Wavelength Covereage} & \colhead{Instrument} & 
\colhead{Resolving Power} & \colhead{Reference} & \colhead{Use}}
\startdata
0.5 - 1.023 $\mu$m & HST STIS & R$\sim$500 & \citet{Schultz1998} & SED\\
1.024 - 2.52 $\mu$m & CGS4 on UKIRT & R$\sim$390-780x$\lambda$ & \citet{Geballe1996} & Retrieval, SED\\
2.98 - 4.15 $\mu$m & NIRC on Keck I & R$\sim$150 & \citet{Oppenheimer1998} & Retrieval, SED\\
4.5 - 5.1 $\mu$m & CGS4 on UKIRT & R$\sim$400x$\lambda$ & \citet{Noll1997} & Retrieval, SED\\
\enddata
\end{deluxetable*}

\section{Literature Data on Gliese 229 System} \label{sec:literature}
\subsection{Discovery, Observations and Mass Controversy of Gl 229B}

The methane-bearing object that confirmed the existence of brown dwarfs in 1995 has been the subject of intense study and debate since its discovery. Separated from its primary at a distance of 7.78 $\pm$ 0.1 arcsec, Gl 229B was initially found as a proper motion companion to an M1V star, Gl 229A \citep{Oppenheimer1995, Nakajima1995}. Initial observations of Gl 229B were done by \citet{Oppenheimer1995} who obtained a low-resolution near-infrared spectrum (1.0-2.5 $\mu$m) on the Hale 200-inch telescope later followed up by higher resolution spectroscopic observations that would cover 0.8-5.0 $\mu$m \citep{Geballe1996, Schultz1998, Saumon2000}. These spectroscopic studies confirmed the presence of H$_2$O and CH$_4$ as well as CO in excess of the abundance predicted by chemical equilibrium. Broadband photometric observations obtained by \citet{Matthews1996, Golimowski1998, Leggett1999, Golimowski2004} combined with a Hipparcos parallax for Gl 229A \citep{Perryman1997} -- since updated in \citet{Gaia2021} -- resulted in a robust determination of its bolometric luminosity.

Several works have attempted to reproduce Gl 229B's spectra using thermo-chemical equilibrium models and subsequently derive its fundamental parameters to varied results. One initial spectroscopic study used PHOENIX grid models \citep{Allard1996} to place upper limits on the effective temperature at 1000 K in an attempt to constrain the mass. However, uncertainties in the age of this system cause a significant challenge in determining fundamental parameters, such as mass, across several forward modelling attempts. As spectroscopic and photometric observations improved, \citet{Saumon2000} employed evolutionary models of \citet{Burrows1997} but were still not able to constrain log(g) better than 5.0 $\pm$ 0.5, although they predicted an atmosphere depleted in heavy metals, particularly oxygen. More recent work by \citet{Nakajima2015} attempted to use measured abundances and an age estimate of Gl 229A to determine an approximately solar metallicity for Gl 229B, assuming co-evality of the pair. \citet{Filippazzo2015} derives fundamental parameters based on the evolutionary models of \citet{Baraffe2003, Saumon2008} to create a distance-calibrated SED and directly integrated bolometric luminosity ($L_{\rm Bol}$). A summary of fundamental parameters for Gl 229B can be found in Table \ref{tab:Gl229B Literature}.

While uncertainty in age estimates for the system posed a challenge in spectroscopic modelling done shortly after its discovery, a dynamical mass measurement for Gl 229B highlights this tension even further. \citet{Brandt2021} used astrometry from \citet{Gaia2021} to constrain a dynamical mass at 71.4 $\pm$ 0.6 $M_{\rm Jup}$, placing Gl 229B at the edge of the stellar mass boundary. All evolutionary model predictions prior to this work estimate an upper limit on age to be $<$ 5 Gyr based on its bolometric luminosity and derived effective temperature with one work even suggesting Gl 229B is as young as 16 Myr \citep{Leggett2002a}. A predicted mass as high as that reported in \citet{Brandt2020, Brandt2021} acutely challenges this age estimate since an object this massive would take much longer to cool to its reported luminosity and approximate temperature of 900 K. In fact, Gl 229B is on a growing list of T dwarfs whose high masses conflict with what one might expect from evolutionary models \citep[for example, $\epsilon$ Indi AB;][]{Dieterich2018}. It is unclear at this point whether this is due to unresolved issues in the models or observational biases. However, in this work, we attempt to derive best-fit fundamental parameters for Gl 229B as well as investigate the plausibility of an anomalously high mass. 

\begin{figure*}[ht!]
\centering
\plotone{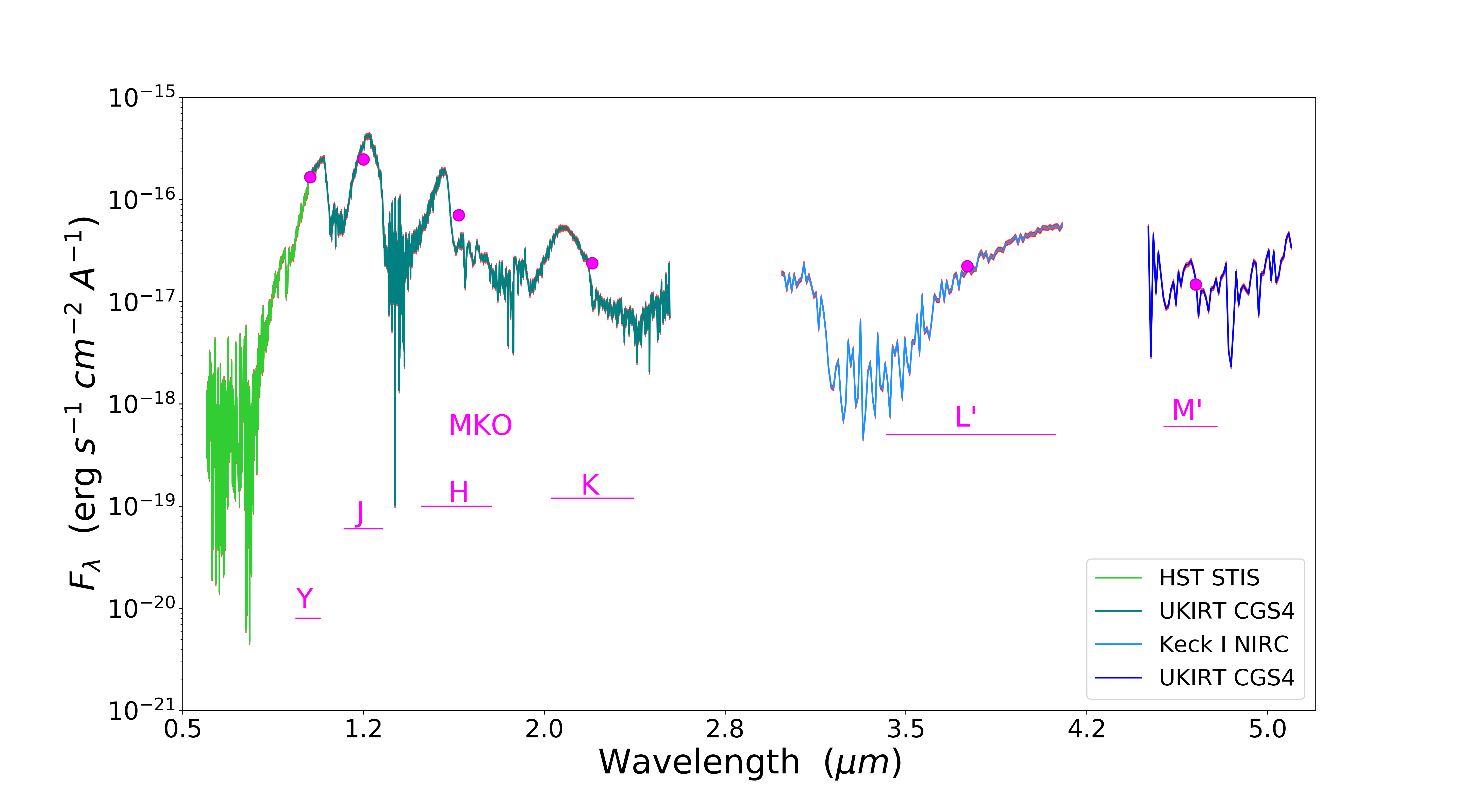}
\caption{Distanced-calibrated SED for Gl 229B using the spectra listed in Table \ref{tab:retrieval data} and photometry listed in Table \ref{tab:SED results}. Spectrum calibration was done using the techniques of \citet{Filippazzo2015}, updated with a system parallax measurement from \citet{Gaia2021}. Pink lines show the wavelength coverage of each photometric band and photometric points are each plotted at their respective band center. Underlying red shows the flux uncertainty which was calculated using SNR=10. \label{fig:SED}}
\end{figure*}

\subsection{Details on the Primary Gl 229A}
In 1995, \citet{Nakajima1995} observed Gl 229A, a known M1V dwarf \citep{Kirkpatrick1991, Cushing2006}, with the Adaptive Optics Coronograph at the Palomar 60-inch telescope in search of a proper motion companion. Gl 229A was initially chosen as a target of interest in a search for brown dwarf companions to stars within 15 pc of the Sun. Its known low space motion \citep{Leggett1992} as well as a precisely measured distance in Hipparcos \citep{Perryman1997} made it a promising candidate. Since the discovery of Gl 229B, Gl 229A has also been a target of several observational studies in an attempt to further our understanding of the physics of low mass stars and the fundamentals of this system in particular. However, there is still seemingly discordant conclusions on its age and chemical composition which have added to the mystery of this system as a whole.

Based on its kinematics and coronal activity, it was originally believed to be younger than the Sun but at least as old as the Hyades ($\sim$0.8 Gyr), about 0.5-5 Gyr \citep{Nakajima1995}. However, M dwarf ages are notoriously difficult to determine due to the long activity lifetimes of fully convective stars \citep{West2008}. Gl 229A is a prime example of this dilemma as later studies are in complete disagreement over its age. \citet{Leggett2002a} used evolutionary models from \citet{Allard2001} to argue that its kinematics and flare star designation support a much younger age of 16-45 Myr. While, more recently, \citet{Brandt2020} used chromospheric and coronal activity along with gyrochronology relations from \citet{Angus2019} to estimate an age of 2.6 $\pm$ 0.5 Gyr. This is further complicated as \citet{Brandt2020} reports a system age of 7-10 Gyr based on the dynamical mass measurement of Gl 229B. They also calculate an age based on Gl 229A's membership in the kinematic thin disk but note that the resulting young age is strongly disfavored by the low levels of chromospheric and coronal activity.

In better agreement among different models are metallicity measurements for Gl 229A. Using medium and high resolution spectra, \citet{Nakajima2015, Gaidos2014, Neves2014, Mould1978, Schiavon1997} find Gl 229A to be consistent with solar values. There is one exception among the literature which found a best fitting model with [M/H] = -0.5 \citep{Leggett2002a}. However, spectroscopic measurements do not seem to support such a low metallicity. Reported metallicity measurements for Gl 229A are listed in Table \ref{tab:Gl229A Literature}. Additionally, \citet{Nakajima2015} reported a C/O ratio for Gl 229A based off of inferred bulk carbon and oxygen abundances published in \citet{Tsuji2014} and \citet{Tsuji2015} and found a slightly supersolar value of 0.68 $\pm$ 0.12 (as compared to a solar value of 0.55 from \cite{Asplund2009}), making it the only determined C/O ratio for Gl 229A.

\subsection{Data Used in Retrievals}
For the purposes of this retrieval work, we began using near infrared data obtained using CGS4 at the United Kingdom 3.8 m Infrared Telescope (UKIRT) \citep{Geballe1996} whose spectra was updated with improved photometry by \citet{Leggett1999}.

To further constrain retrieved chemical abundances and other fundamental parameters, L band spectral data from \citet{Oppenheimer1998} and M band spectral data from \citet{Noll1997} were included, making a combined spectral wavelength coverage 1.0-5.0 $\mu$m. Table \ref{tab:retrieval data} lists the spectra used in this work.

\section{Results from the SED} \label{sec:SED}
We re-visit the SED in Figure \ref{fig:SED} -- done previously in \citet{Filippazzo2015} -- using the parallax of Gl 229A from the latest Gaia data release, Gaia EDR3 \citep{Gaia2021}. As in \citet{Filippazzo2015}, both the optical and the infrared spectrum were used to construct the SED. The SED is scaled and distance-calibrated with the available photometry and EDR3 parallax which we use to find a distance of 5.761 $\pm$ 0.001 pc. We follow the exact same methodology as \citet{Filippazzo2015}, only updating the SED with this new parallax. The spectra used in the SED are listed in Table \ref{tab:retrieval data} while the parallax and photometry are listed in Table \ref{tab:SED results}. The original data did not include a flux uncertainty array so we chose a conservative estimate of SNR=10 for the entire spectrum.

Bolometric luminosity ($L_{\rm bol}$) was calculated by integrating under the distance-calibrated SED from 0 - 1000 $\mu$m. To account for wavelength coverage gaps in the spectra, the flux in each region was estimated by linearly interpolating to zero from the short wavelength limit and appending a Rayleigh-Jeans tail at the long wavelength limit \citep{Filippazzo2015}. The radius and mass estimates are calculated by comparison to evolutionary models using the derived $L_{\rm bol}$ value and object age. We conservatively assume an object age between 0.5 - 10 Gyr since there are no obvious spectral signatures of youth. The range of predicted radius and mass values comes from both the cloudy and cloudless evolutionary models of \citet{Saumon2008} and \citet{Baraffe2003}. The range of values were taken to be the minimum and maximum from all model predictions. The effective temperature, $T_{\rm eff}$, was calculated by the Stefan-Boltzmann law, using the predicted radius and derived $L_{\rm bol}$. For a more detailed explanation of the methodology used in the construction of this SED or its derived parameters, see \citet{Filippazzo2015}. Fundamental parameters derived from the SED are listed in Table \ref{tab:SED results} as well as Table \ref{tab:Gl229B Literature}.

\begin{deluxetable*}{lr}
\tablenum{3}
\tablewidth{25pt}
\tablecaption{Priors for Gl 229B Retrieval Models}
\tablehead{
\colhead{Parameter} & \colhead{Prior}}
\startdata
gas volume mixing ratio & uniform, log $f_{gas} \geq$ -14, $\Sigma_{gas} f_{gas} \leq$ 1\\
thermal profile ($T_{bottom}$, $T_{top}$, $T_{middle}$, $T_{q1}$, $T_{q3}$) & uniform, 0 K $<$ T $<$ 4000 K\\
radius & uniform, 0.5$R_{\rm Jup}$  $\leq$ R $\leq$ 2.0$R_{\rm Jup}$ \\
mass\tablenotemark{a} & uniform, 1$M_{\rm Jup}$  $\leq$ M $\leq$ 80$M_{\rm Jup}$ \\
cloud top\tablenotemark{b} & uniform, -4 $\leq$ logP$_{CT}$ $\leq$ +2.3\\
cloud decay scale\tablenotemark{c} & uniform, 0 $<$ log$\Delta$P$_{decay}$ $<$ 7\\
cloud thickness\tablenotemark{d} & uniform, logP$_{CT}$ $\leq$ log(P$_{CT}$ + $\Delta$P) $\leq$ 2.3\\
cloud total optical depth (extinction) & uniform, 0 $\leq$ $\tau_{cloud}$ $\leq$ 100\\
single scattering albedo & constant, $\omega_0$ = 0\\
wavelength shift & uniform, -0.01 $<$ $\Delta\lambda$ $<$ 0.01 $\mu$m\\
tolerance factor & uniform, log(0.01 x $min(\sigma_i^2)$) $\leq$ b $\leq$ log(100 x $max(\sigma_i^2$))\\
\enddata
\tablenotetext{a}{This mass prior range was constrained to 70-72 $M_{\rm Jup}$  for a single, cloudless model. See Section \ref{sec:best fit mass}.}
\tablenotetext{\tiny b}{For a deck cloud, this is the pressure where $\tau_{cloud}$=1, for a slab cloud this is the top of the slab.}\vspace{-2ex}
\tablenotetext{\tiny c}{Decay height for cloud deck above the $\tau_{cloud}$=1.0 level.}\vspace{-2ex}
\tablenotetext{\tiny d}{Thickness and $\tau_{cloud}$ retrieved only for slab cloud.}\vspace{-3ex}
\label{tab:priors}
\end{deluxetable*}

\section{Brewster Framework} \label{sec:brewster}
The retrieval models presented here were constructed with the \textit{Brewster} retrieval framework \citep{Burningham2017, Burningham2021}. In this section, we provide a brief summary of \textit{Brewster} as well as any modifications made for this work. We differ from \citet{Burningham2017, Burningham2021} with the use of a version update that utilizes nested sampling via \texttt{PyMultiNest} \citep{Buchner2014} instead of \texttt{EMCEE} \citep{Foreman2013}. For a more detailed description of this framework with a focus on the \texttt{EMCEE} sampler, see \citet{Burningham2017}.

\subsection{Forward Model}
The forward model consists of the radiative transfer solver, thermal profile, and opacity and scattering properties as a function of wavelength. The forward model solves for emergent flux from radiative transfer using the two stream technique of \citet{Toon1989}. This includes scattering, first introduced by \citet{McKay1989} and later used by \citet{Marley1996}, \citet{Saumon2008} and \citet{Morley2012}. We use a 64 pressure layer atmosphere (65 levels) with geometric mean pressures in range -4 $<$ log P $<$ 2.3 in bars, spaced at 0.1 dex intervals.

\subsubsection{Thermal Profile}
 The thermal profile used is a computationally simple five point parameterization in which we specify five temperature-pressure points: the top ($T_{top}$), bottom ($T_{bottom}$) and middle of the atmosphere ($T_{middle}$) and two midpoints between the top and middle ($T_{q1}$), and bottom and middle ($T_{q3}$). These points are calculated in order beginning with $T_{bottom}$ which is selected in range between zero and the maximum temperature defined in our prior. This work defines a maximum temperature of 4000 K. Then $T_{top}$ is chosen between zero and $T_{bottom}$, $T_{middle}$ chosen between $T_{top}$ and $T_{bottom}$, and the remaining two midpoints chosen between $T_{top}$ and $T_{middle}$ ($T_{q1}$), $T_{middle}$ and $T_{bottom}$ ($T_{q3}$). A uniform prior is assumed for each temperature within its respective range. This does not allow for temperature inversions but can result in ``wobbly" profiles.

\subsubsection{Gas Opacities}
For the models presented in this work, we assume uniform-with-altitude mixing ratios for absorbing gases and calculate layer optical depths using high-resolution (R=10,000) opacities from \citet{Freedman2008, Freedman2014}.

Particularly important in the cooler atmospheres of brown dwarfs are the D resonance doublets of Na-I ($\sim$ 0.59 $\mu$m) and K-I ($\sim$ 0.77 $\mu$m) that create a defining spectral feature in the range 0.4 - 1.0 $\mu$m. In T dwarfs, these line profiles can be detected up to $\sim$ 3,000 $cm^{-1}$ from the line center \citep[e.g.][]{Burrows2000, Liebert2000, Marley2002, King2010}, making the Lorentzian line profile insufficient. Instead, we implement line wing profiles based on the unified line shape theory \citep{Allard2007a, Allard2007b}. In previous retrieval work with T dwarfs, \citet{Line2017} suggested the use of alkali opacities from \citet{Burrows2003} that calculate absorption line profiles for the D1 and D2 lines of Na-I and K-I broadened by H$_2$ - and He collisions for effective temperatures below 2000K and perturber densities derived from the quasi-static theory of absorption \citep{Holtsmark1925, Holstein1950}. However, \citet{Gonzales2020} showed that there may not be a clear distinction in preferred tabulated line profiles for T dwarfs between those from \citet{Burrows2003} and Allard N. (private communication). Broadened D1 and D2 line profiles from Allard N. are calculated for temperatures in the range 500 - 3000 K and perturber densities up to $10^{20}$ $cm^{-3}$ where two collisional geometries are considered for broadening by H$_2$. Profiles within 20 $cm^{-1}$ of the line center are Lorentzian with a width calculated from the same theory.

Line opacities are tabulated across the temperature-pressure regime in 0.5 dex steps for pressure and in steps from 20K to 500K for the temperature range 75K - 4000K. This is then linearly interpolated to our working pressure grid. We also include Rayleigh scattering for H$_2$, He, and CH$_4$ only. We assume atmospheric proportions of 0.84 H$_2$ + 0.16 He based on Solar abundances to give an effective broadening width for each line. After retrieving abundances of the gases assumed to be in the atmosphere, neutral H, H$_2$ and He are assumed to make up the remainder of the gas in a layer.

\subsubsection{Gas Abundances}
As stated in the previous section, we assume uniform-with-altitude mixing ratios, as opposed to layer-by-layer varying gas mixing ratios, for all absorbing gases and retrieve the overall abundances directly for all models in this work. It should be noted that the uniform-with-altitude mixing method is a simplification in our model that cannot distinguish variations in gas abundance with altitude for some species, particularly the alkalies which can vary by several orders of magnitude in the photosphere or CO which is believed to be in chemical disequilibrium in the photosphere \citep{Fegley1996, Oppenheimer1998}. However, we utilize the uniform-with-altitude method because the layer-by-layer approach is computationally prohibitive.

\begin{deluxetable*}{llcc}[ht]
\tablenum{4}
\tablecaption{Model Selection for Gliese 229B}
\tablewidth{25pt}
\tablehead{\colhead{Model} & \colhead{Alkali} & \colhead{Number of Parameters} & \colhead{$\Delta$logEvidence}}
    \startdata
    Cloudless & Allard & 18 & 0\\
    Grey Deck Cloud & Allard & 20 & 8.78\\
    Grey Slab Cloud & Allard & 21 & 12.72\\
    Power Law Deck Cloud & Allard & 21 & 13.53\\
    ZnS Deck Cloud & Allard & 22 & 14.33\\
    KCl Slab Cloud & Allard & 23 & 17.61\\
    KCl Deck Cloud & Allard & 22 & 19.41\\
    Power Law Slab Cloud & Allard & 22 & 23.78\\
    ZnS Slab Cloud & Allard & 23 & 30.10\\
    Cloudless & Burrows & 18 & 32.21\\
    Cloudless, Mass Prior & Allard & 17 & -0.85\tablenotemark{a}\\
    \enddata
\tablenotetext{a}{Putting a constraint on mass effectively removes one parameter from the model by increasing model confidence. This is why we see a relative increase in logEv for this model compared to our best fit and why we cannot directly compare this model against any other we have tested. See Section \ref{sec:best fit}.}
\label{tab:Retrieval models}
\end{deluxetable*}

\subsubsection{Cloud Modelling}
The cloud parameterizations we utilize in this work closely follow the methodology described in \citet{Burningham2017} and \citet{Gonzales2020, Gonzales2021}. As in \citet{Burningham2017, Burningham2021}, we define two categories of clouds, “slab" and “deck". Both slab and deck clouds have an opacity distributed among layers in pressure space and an optical depth determined by cloud designation as either grey or non-grey. Total optical depth ($\tau_{cloud}$) for a grey cloud model is calculated at 1$\mu$m. For a non-grey cloud application, we use a power-law distribution to describe the optical depth, $\tau = \tau_0\lambda^{\alpha}$, where $\tau_0$ is the optical depth at 1$\mu$m. For the case of a non-grey cloud, we designate another model parameter, the power ($\alpha$) in the optical depth. We discuss a cloud's optical depth in terms of extinction and assume an absorbing cloud by setting the single scattering albedo to zero, as done in \citet{Gonzales2020, Gonzales2021}.

Beyond the grey and non-grey cloud parameterizations, we follow the work of \citet{Burningham2021} by testing different condensate species under the assumption of Mie scattering. In particular, we investigate the impact of zinc sulfide (ZnS) and potassium chloride (KCl) condensates using refractive indices from \citet{Wakeford2015} and pre-tabulated Mie coefficients as a function of particle radius and wavelength. Wavelength-dependant optical depths and phase angles in each pressure layer are calculated by integrating the cross-sections and Mie efficiencies over the particle size distribution in that layer for a given cloud species. Particle number density in a layer is calibrated to the optical depth at 1 $\mu$m as determined by the cloud type (slab or deck). We assume either a Hansen (\citeyear{Hansen1971}) or lognormal distribution of particle sizes. The Hansen distribution for particle number $n$ with radius $r$ is given:
\begin{equation}
    n(r) \propto r^{\frac{1-3b}{b}}e^{-\frac{r}{ab}}
\end{equation}
where $a$ and $b$ are the effective radius and spread of the distribution, respectively. These parameters are defined as:
\begin{equation}
    a = \frac{\int_0^{\infty}r\pi r^2n(r)dr}{\int_0^{\infty}\pi r^2n(r)dr}
\end{equation}
\begin{equation}
    b = \frac{\int_0^{\infty} (r-a)^2\pi r^2 n(r)dr}{a^2 \int_0^{\infty} \pi r^2n(r)dr}
\end{equation}

A deck cloud is parameterized by the cloud top pressure, P$_{top}$, the decay height, $\Delta log P$, and the cloud particle single scattering albedo which we set to zero. The cloud top pressure is defined as the point in pressure space at which the optical depth passes unity, or $\tau$ = 1, (looking down). The decay scale pressure describes how the optical depth changes with changing pressure from the cloud deck and is defined as $d\tau / dP \propto exp((P - P_{deck})/\Phi$), where P$_{deck}$ is the height at which the cloud is optically thick and $\Phi$ = P$_{top}$(10$^{\Delta logP}$-1)/10$^{\Delta logP}$ is the decay scale of the cloud in bars. The deck cloud becomes optically thick at P$_{top}$ so for P $>$ P$_{top}$, optical depth increases following the decay function until it reaches $\Delta\tau_{layer}$ = 100. The deck cloud becomes opaque with increasing pressure relatively quickly as a result of the decay function so we obtain little atmospheric information from deep below the cloud top. To account for this, the pressure-temperature (P-T) profile below the cloud deck is an extension of the gradient and spread at the cloud top pressure.

The distinguishing marker for a slab cloud is the addition of another parameter to determine the total optical depth at 1 $\mu$m ($\tau_{cloud}$), since we can “see" the bottom of this type of cloud. The optical depth is distributed throughout the cloud as $d\tau $/$ dP$ $\propto$ P (looking down), where its maximum value is reached at the bottom of the slab (highest pressure). In principle, the slab can have any optical depth but we restrict the prior to 0.0 $\leq \tau_{cloud} \leq$ 100.0. Instead of considering the decay scale pressure, as we did for a deck cloud, we consider the cloud thickness in $d$logP and parameterize for cloud top pressure, P$_{top}$.

\subsection{Retrieval Model}
The retrieval process depends on the chosen elements of the parameter set. Changing the elements that are passed to the forward model have effects on the resultant spectrum. Optimizing the forward model's fit to the data by varying the parameter set, or state-vector, takes place within a Bayesian framework. A detailed explanation of this framework can be found in \citet{Burningham2017}. To summarize, \textit{Brewster} applies Bayes' theorem to calculate the “posterier probability", $p(\textbf{x}|\textbf{y})$, the probability of a set of parameters' $(\textbf{x})$ truth value given some data $(\textbf{y})$, in the following way:
\begin{equation}
    p(\textbf{x}|\textbf{y}) = \frac{\mathscr{L}(\textbf{x}|\textbf{y})p(\textbf{x})}{p(\textbf{y})}
\end{equation}
where $\mathscr{L}(\textbf{x}|\textbf{y})$ is the likelihood that quantifies how well the data match the model, $p(\textbf{x})$ is the prior probability on the parameter set and $p(\textbf{y})$ is the probability of the data marginalized over all parameter values, also known as the Bayesian evidence. As detailed in \citet{Burningham2017}, the original version of \textit{Brewster} uses \texttt{EMCEE} \citep{Foreman2013} to sample posterior probabilities and inflate errors using a tolerance parameter to allow for unaccounted sources of uncertainty. The \texttt{EMCEE} chain requires tens of thousands of iterations over hundreds of parallel walkers in order to converge and can often require several days of computing to complete a single round of parameter estimation depending on wavelength range and model complexity. Additionally, the use of \texttt{EMCEE} as a sampler can result in models with degenerate parameter solutions or local, rather than global, maximum likelihoods.

In this new instance of \textit{Brewster}, the posterior probability space is explored using the \texttt{PyMultiNest} sampler \citep{Buchner2014, Feroz2011} which utilizes nested sampling to discover the set of parameters with the maximum likelihood given the data. \texttt{PyMultiNest} is a Bayesian inference tool that explores the parameter space to maximize the likelihood of the forward model fit to the data. The samples in an n-dimensional hypercube, or state-vector, are translated into parameter values via the “prior-map". The prior-map function is how prior probabilities are set for each parameter and are then transformed into appropriate parameter values to be used in the forward model. This algorithm is equipped to handle a parameter space that may contain multiple posterior modes and/or degeneracies in moderately high dimensions.

Unlike the procedure in \citet{Burningham2017}, we retrieve radius and mass directly which are then used to numerically determine a value for gravity. The radius is determined from the scaling factor required to match the absolute flux from the forward model to the data and the measured parallax. The radius is restricted to be within the range 0.5 - 2.0 $R_{\rm Jup}$  and the mass between 1 - 80 $M_{\rm Jup}$  as suggested by \citet{Saumon2008}, COND \citep{Baraffe2003} and DUSTY \citep{Chabrier2000, Baraffe2002} substellar evolutionary models. In order to investigate a dynamical mass measurement of 71.4 $\pm$ 0.6 $M_{\rm Jup}$  \citep{Brandt2021}, we restrict the probability space for mass to be 70-72 $M_{\rm Jup}$   in just one of our model investigations. Table \ref{tab:priors} lists the priors used in our modelling which are based on those defined in \citet{Burningham2017}.

We began this modelling by investigating the impact of running a retrieval model on near-infrared data only (1.0-2.5 $\mu$m) as opposed to the entire combined infrared spectrum. We are unable to constrain CO when using only NIR data, which is known to be in excess of thermo-chemical equilibrium abundance in the photosphere \citep{Oppenheimer1998}. This result is consistent with the work of \citet{Line2015, Line2017} which could not constrain CO, CO$_2$ or H$_2$S with NIR data, alone. The first observable CO feature in a T dwarf spectrum is expected to be the first vibration-rotation band (1-0) at 4.7 $\mu$m which was observationally confirmed in Gl 229B by \citet{Noll1997} and then again by \citet{Oppenheimer1998}. Therefore, with the addition of extended infrared data (2.98-5.0 $\mu$m), we are able to constrain CO abundance across models.

For all models we use the distance-calibrated (to 10 pc) SED beginning at 1.0 $\mu$m and out to 5.0 $\mu$m. This spectrum calibration differs from the method used in \citet{Burningham2017} in which they calibrated spectra to the 2MASS J-band photometry and used the object's true distance in their initialization.

We retrieve the following gases known to sculpt T dwarf spectra: H$_2$O, CO, CH$_4$, NH$_3$, Na, and K. We tie Na and K together as a single element in the state vector assuming a Solar ratio taken from \citet{Asplund2009} \citep{Line2015, Burningham2017, Gonzales2020, Gonzales2021}. We are consistent with \citet{Gonzales2020} in excluding CO$_2$ and H$_2$S from our gas list as we are still unable to constrain these even with the inclusion of longer wavelength data. We test multiple cloud parameterizations, beginning with the most simple, cloudless model and building to a 5 parameter Mie scattering cloud slab model.

\subsection{Model Selection Parameters}
Across models, we focused on making changes in our approach to cloud parameterization while holding fixed the gases included in each model, as well as gas abundance method and alkali opacity. In order to compare these different models, we use a calculation of the Bayesian evidence, specifically the logEvidence or logEv, where the highest logEv is preferred. We use the following selection criterion from \citet{Kass1995} to distinguish between two models, with evidence against the lower logEv as:
\begin{itemize}
    \item 0 $<$ $\Delta$logEv $<$ 0.5: no preference worth mentioning
    \item 0.5 $<$ $\Delta$logEv $<$ 1: positive
    \item 1 $<$ $\Delta$logEv $<$ 2: strong
    \item $\Delta$logEv $>$ 2: very strong
\end{itemize}
We began by building from the least complex model (cloud-free) to the most complex slab cloud model. We compare Burrows and Allard alkali opacities for the best fit model as there is conflicting evidence over which is preferred for T dwarf retrievals \citep{Line2017, Gonzales2020}.

\begin{deluxetable}{lrcr}
\tablenum{5}
\tabletypesize{\footnotesize}
\tablecaption{Cloudless, Allard Alkali Model Parameters}
\tablewidth{25pt}
\tablehead{\colhead{Parameter} & \colhead{} & \colhead{Value} & \colhead{}}
    \startdata
     Model & Best Fit\tablenotemark{a} &  & Mass Constrained\tablenotemark{b}\\
     \hline
     \textbf{Retrieved} & & \\
    \hline
    $H_2O$ & $-3.53 \pm 0.04 $ & & $-3.49 \pm 0.03$\\
    CO & $-4.59^{+0.21}_{-0.24}$ & & $-4.56^{+0.22}_{-0.25}$\\
    $CH_4$ & $-3.38 \pm 0.05$ & & $-3.32 \pm 0.03$\\
    $NH_3$ & $-4.58 \pm 0.05$ & & $-4.51 \pm 0.03$\\
    Na+K & $-5.69 \pm 0.03 $ & & $-5.67 \pm 0.03$ \\
    Radius ($R_{\rm Jup}$) & $1.10 \pm 0.04$ & & $1.12 \pm 0.04$\\
    Mass ($M_{\rm Jup}$) & $50^{+12}_{-9}$ & & $71 \pm 1$\\
    \hline
    \textbf{Derived} & & \\
    \hline
    log g (dex) & $5.01^{+0.10}_{-0.09}$ & & $5.15 \pm 0.03$\\
    $T_{\rm eff}$ (K) & $834^{+17}_{-15}$ & & $829^{+19}_{-17}$\\
    log($L_{\rm Bol}$ /$L_{\rm Sun}$) & $-5.25^{+0.02}_{-0.01}$ & & $-5.25^{+0.02}_{-0.01}$\\
    C/O & $1.38^{+0.08}_{-0.07}$ & & $1.45 \pm 0.07$\\
    C/O\tablenotemark{c} & $1.06^{+0.06}_{-0.05}$ & & $1.11^{+0.06}_{-0.05}$\\
    $[M/H]$ & $-0.24 \pm 0.05$ & & $-0.19 \pm 0.03$\\
    $[C/H]$ & $-0.01 \pm 0.05$ & & $0.05 \pm 0.03$\\
    $[O/H]$ & $-0.41^{+0.05}_{-0.04}$ & & $-0.37 \pm 0.04$\\
    \enddata
\tablenotetext{\tiny a}{Best fit model without any prior constraints placed on retrieved parameters.}\vspace{-1ex}
\tablenotetext{\tiny b}{Best fit model with a prior constraint placed on the mass.}\vspace{-1ex}
\tablenotetext{\tiny c}{Oxygen-corrected C/O ratio (see Section \ref{sec:chemistry})}\vspace{-2ex}
\label{tab:Retrieval results}
\end{deluxetable}

\section{Retrieval of Gl 229B}\label{sec:retrieval}
Table \ref{tab:Retrieval models} lists all tested models as well as their $\Delta$logEv relative to the best fit model: the cloudless, Allard alkali model. Based on our selection criterion, it is clear that all models including clouds were strongly rejected, all with $\Delta$logEv greater than 2, suggesting that no cloud model can provide a strong fit to the spectroscopic features observed in Gl 229B. Additionally, we show that a cloudless model using Burrows alkalies is also strongly rejected as compared to cloudless with Allard alkalies. We also list the cloudless model with a mass prior which shows a $\Delta$logEv that is positively preferred over our best fit cloudless model. However, since we tightly restrict the mass range for this one model, we effectively remove one degree of freedom and expect model confidence to increase as a result. Therefore, we cannot directly compare our mass constrained model against any other models we have tested here.

\begin{figure*}[ht!]
\centering
\plotone{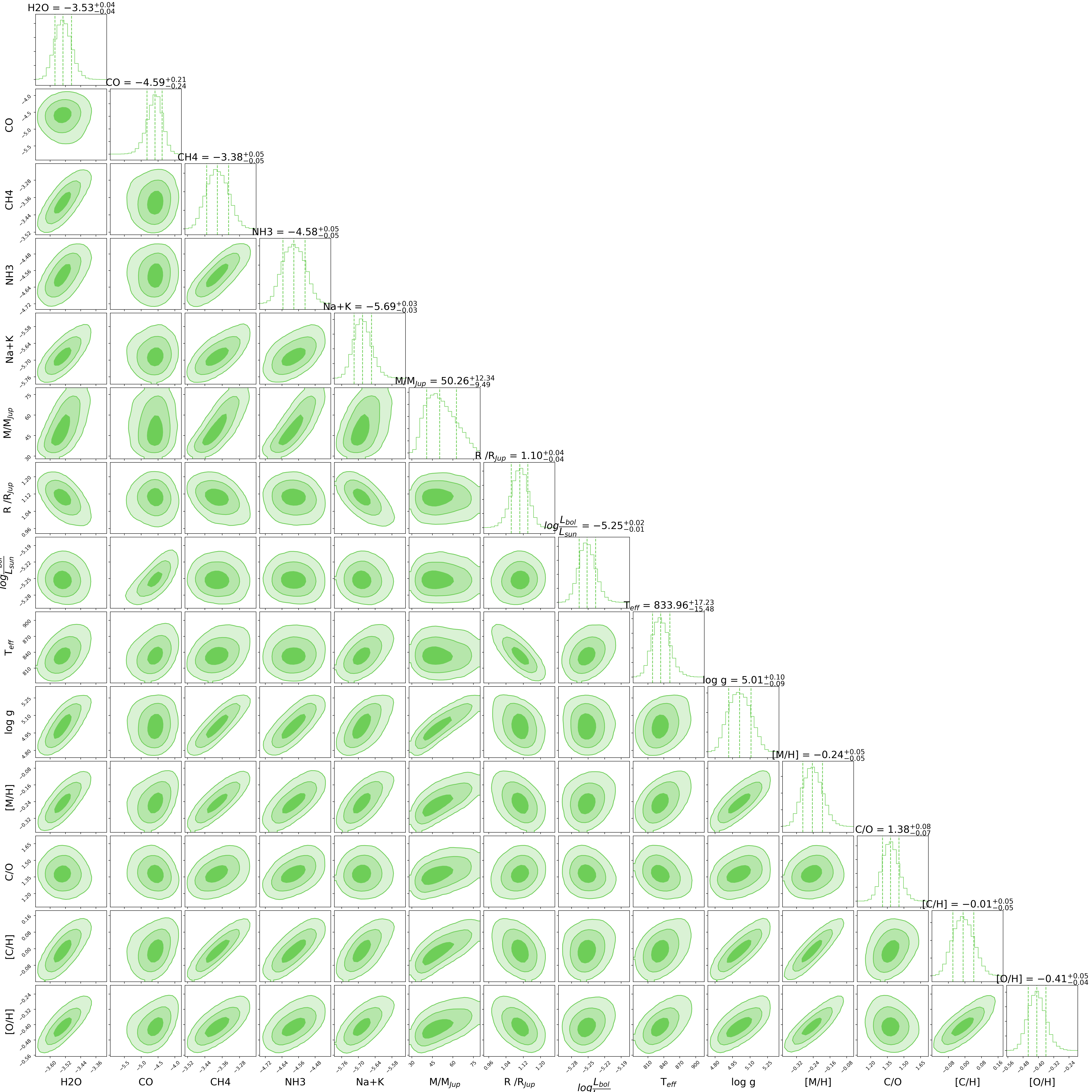}
\caption{\textit{Best Fit Model: Cloudless, Allard Alkalies}
 Posterior probability distributions for retrieved gas abundances, mass, radius and derived quantities of the best fit cloudless model with Allard alkalies. Far right diagonal plots show marginalized posteriors for each parameter along with 2D parameter correlation histograms. Dashed lines show the median retrieved value (reported value) along with 1$\sigma$ confidence intervals. Gas abundances are in units of dex. \label{fig:corner}}
\end{figure*}

\begin{figure*}[ht]
\centering
\plotone{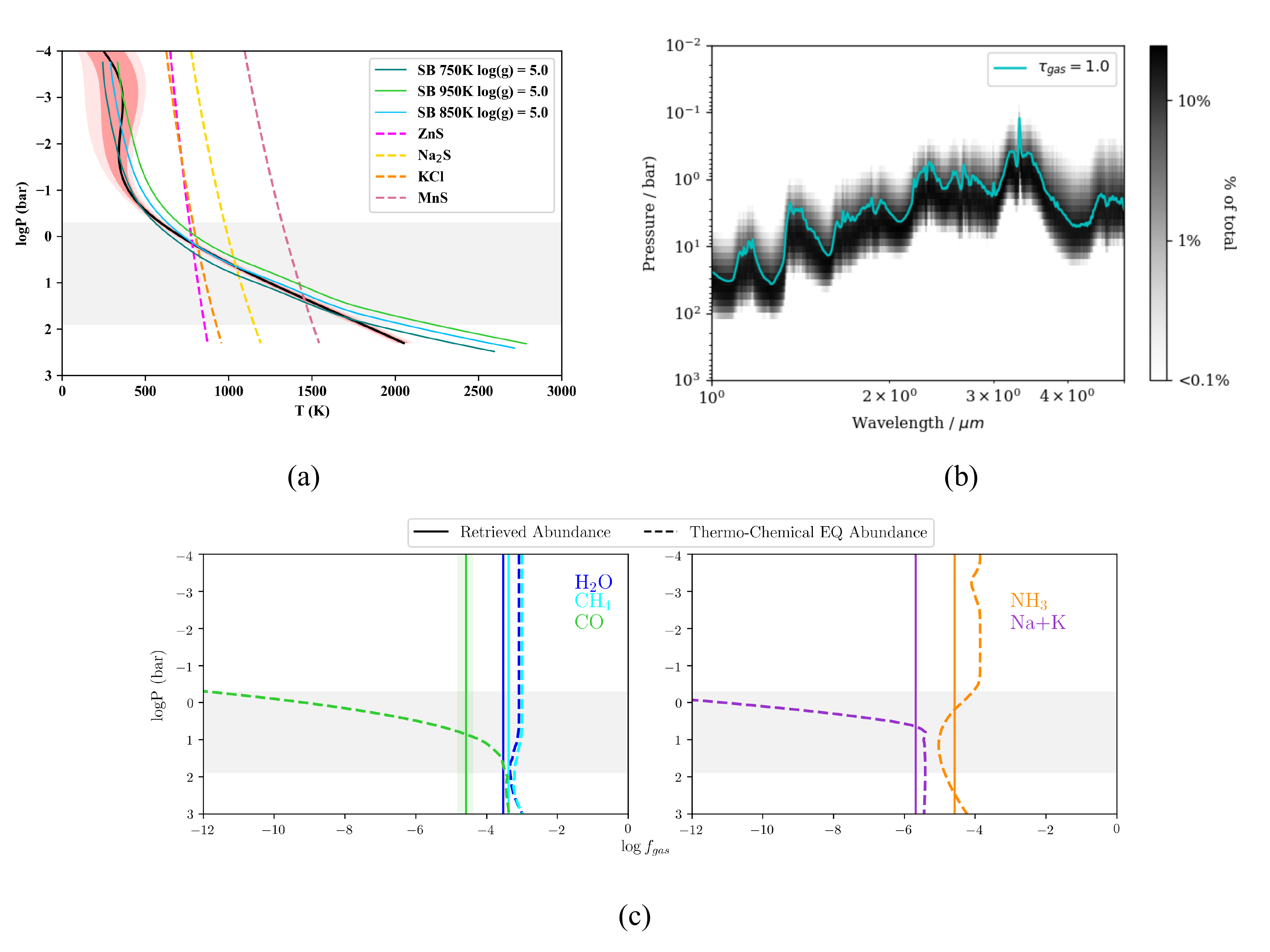}
\caption{\textit{Best Fit Model: Cloudless, Allard Alkalies} Figures on the top line show (a) retrieved temperature-pressure profile and (b) contribution function for the best fit model. Our maximum likelihood retrieved profile is shown in black compared to cloudless Sonora Solar model profiles (green and blue) at similar temperature to our retrieved $T_{\rm eff}$ . Dashed lines show condensation curves for possible cloud species. In the bottom panel (c) we show retrieved uniform-with-altitude mixing abundances as compared to thermo-chemical equilibrium grid abundances for supersolar C/O (1.0 relative to 0.55 of \citet{Asplund2009}) and Solar metallicity. In both figures (a) and (c), we show the approximate location of the photosphere as determined by our contribution function with the inclusion of a light grey panel. \label{fig:profile and contribution}}
\end{figure*}

\subsection{Best Fit Model: A Cloudless Atmosphere}\label{sec:best fit}
\subsubsection{Retrieved Gas Abundances and Fundamental Parameters}
Figure \ref{fig:corner} shows the posterior probability distributions for retrieved gas abundances, mass and radius as well as log(g), $T_{\rm eff}$ , $L_{\rm Bol}$ , C/O ratio, [M/H], [C/H] and [O/H] from our best fit model which are calculated based on retrieved quantities. We list the values from Figure \ref{fig:corner} in Table \ref{tab:Retrieval results} for ease of reading. 

The derived $T_{\rm eff}$  and log(g) are calculated from the retrieved radius and mass along with the parallax measurement. The scale factor ($R^2/D^2$) is calculated from the retrieved radius and parallax. $T_{\rm eff}$  is then determined using this derived scale factor and by integrating the flux in the resultant forward model spectra between 0.6 - 20 $\mu$m. We find that our retrieval-based $T_{\rm eff}$  is cooler than the semi-empirical value found from the SED, although the large uncertainty on the SED generated value allows for 1.2$\sigma$ agreement. However, this $\sim$100 K temperature difference is due to our retrieved radius being $\sim$0.16 $R_{\rm Jup}$  larger than its SED value. Our retrieved radius and mass both agree within 1$\sigma$ to the values determined from evolutionary models when generating the SED. Our derived value for log(g) is also within 1$\sigma$ agreement with its SED value. Due to the uncertainty on mass being relatively large for both our retrieval model and the SED generated value, as well as inconsistency with the reported dynamical mass, we discuss this fundamental parameter in greater detail in Section \ref{sec:mass}. 

The C/O ratio is calculated under the assumption that all of the oxygen exists in H$_2$O and CO and all of the carbon exists in CO and CH$_4$. The following equations were used to derived a value for metallicity, [M/H]:
\begin{subequations}\label{metallicityeqn}
\begin{equation}
    \mathit{f}_{H_2} = 0.84 \times (1 - \mathit{f}_{gas})
\end{equation}
\begin{equation}
    N_H = 2\mathit{f}_{H_2}N_{tot}
\end{equation}
\begin{equation}
    N_{element} = \sum_{molecules} n_{atoms}\mathit{f}_{molecule}N_{tot}
\end{equation}
\begin{equation}
    N_M = \sum_{element} \frac{N_{element}}{N_H}
\end{equation}
\begin{equation}
    [M/H] = log(\frac{N_M}{N_{sun}})
\end{equation}
\end{subequations}
where $\mathit{f}_{H_2}$ is the fraction of $H_2$, $\mathit{f}_{gas}$ is the total gas fraction, $N_H$ is the number of neutral hydrogen atoms, $N_{element}$ is the number of atoms for each element, $n_{atoms}$ is the number of atoms for a given element contained in a single molecule and $N_{tot}$ is the total number of gas molecules. In our calculation of metallicity, $N_{sun}$ is determined using the same formula as $N_M$, using the sum of the solar abundances compared to H. Both [C/H] and [O/H] were calculated using the same procedure for total metallicity using just single element abundances.

As we do not have an SED-based comparison for C/O ratio or metallicity, we discuss these parameters in greater detail in Section \ref{sec:chemistry}.

\begin{figure*}
\centering
\plotone{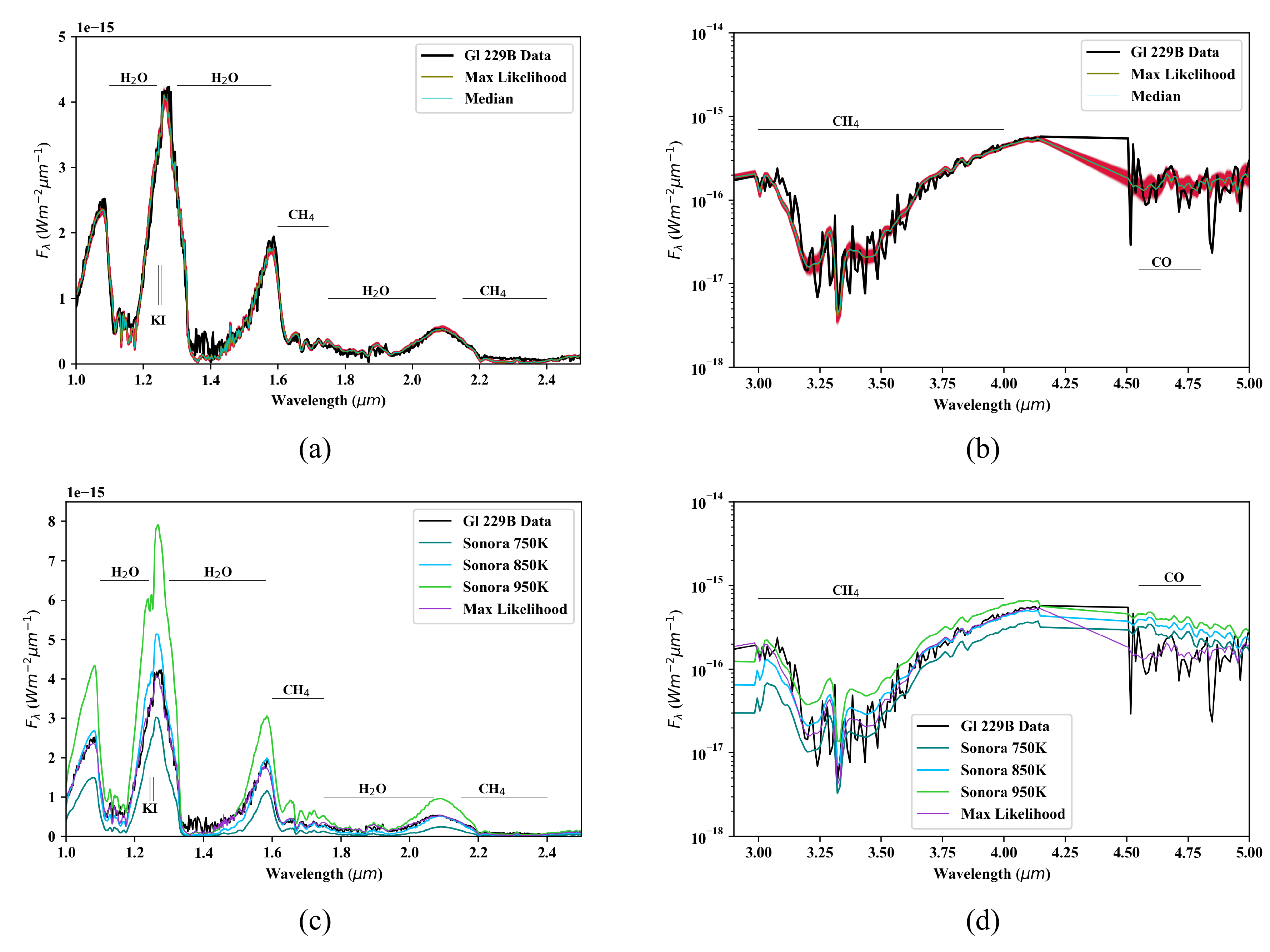}
\caption{\textit{Best Fit Model: Cloudless, Allard Alkalies} On top (a,b) we show the median and maximum likelihood retrieved spectra as compared to observed data for the best model. In red, we show flux uncertainty on this retrieved spectrum. In the bottom panel (c,d), we show the maximum likelihood retrieved spectrum and observed spectrum as compared to Sonora grid models of Solar metallicity and C/O at 750, 850, 950 K (blue and green).
\label{fig:retrieved spectra}}
\end{figure*}

\subsubsection{Temperature-Pressure Profile, Contribution Function and Abundances}
Figure \ref{fig:profile and contribution}(a) shows the retrieved temperature versus pressure (T-P) profile for the best fit cloudless model for Gl 229B. Overplotted in this figure are the Sonora grid models \citep{Marley2021} using Solar-metallicity. The slight thermal inversion apparent in the very top of the atmosphere ($\sim$ 0.01 - 0.001 bar) in our retrieved profile is a computational consequence of our initialized five point profile and not thought to actually exist in that way. As a result, we have much larger 1$\sigma$ and 3$\sigma$ confidence ranges in that pressure space. Our retrieved profile is in best agreement with the Sonora Solar-metallicity log(g) = 5.0, 850K model throughout the photosphere ($\sim$ 0.5 - 20 bar). While the 1$\sigma$ and 3$\sigma$ bounds are relatively small in this pressure range, the log(g) = 5.0, 850 K model fits the profile almost entirely within these bounds. This model agrees to within 3$\sigma$ in the upper atmosphere, as well. Comparing our retrieved profile to the Solar-metallicity log(g) = 5.0, 750 K model, our profile is $\sim$ 50 K warmer throughout most of the photosphere although we do see 1$\sigma$ agreement at the top of the atmosphere ($\leq$ 0.3 bar) as well as the bottom of the photosphere. For the log(g) = 5.0, 950 K model, it is clear that our profile is at least 100 K cooler throughout the photosphere and deeper into the atmosphere.

Figure \ref{fig:profile and contribution}(b) shows the contribution function indicating the location of an optical depth of $\tau$ = 1 for the gas opacity. The contribution function is calculated in each pressure layer in the following way:
\begin{equation}\label{contributionfunc}
    C(\lambda, P) = \frac{B(\lambda, T(P))\int_{P_1}^{P_2} d\tau}{e^{\int_{P_0}^{P_2}d\tau}}
\end{equation}
where B($\lambda$, T(P)) is the Planck function, P$_0$ is the pressure at the top of the atmosphere, P$_2$ is the pressure at the bottom of the layer and P$_1$ is the pressure at the top of the layer. The majority of the flux in the near infrared spans a relatively large portion of the atmosphere from $\sim$ 0.5-80 bar compared to the flux contribution in the early mid infrared that originates higher in the photosphere around $\sim$ 0.2-15 bar.

Figure \ref{fig:profile and contribution}(c) shows retrieved gas abundances as compared to thermo-chemical equilibrium grid model abundances. These chemical grids were calculated using the NASA Gibbs minimization CEA code \citep{Mcbride94} based on prior thermo-chemical models \citep{Fegley1994, Fegley1996,Lodders1999, Lodders2002, Lodders2004, Lodders2010,LodFeg2002, LodFeg2006,Visscher2006,Visscher2010,Visscher2012,Moses2012, Moses2013}. These grids are then used to determine thermo-chemical equilibrium abundances of various atmospheric species for pressure in range 1 microbar - 300 bar and temperatures in range 300 - 4000 K. In this case, we compare against chemical grids of Solar metallicity and supersolar C/O ratio of 1.0 relative to the \citet{Asplund2009} Solar value of 0.55.

Since our model calculates uniform-with-altitude abundances, as opposed to the more physically plausible varying-with-altitude mixing ratios (see Section \ref{sec:brewster}), our retrieved values for CO and Na + K show the median retrieved abundance despite their equilibrium abundances trailing off toward lower values at the top of the photosphere. This retrieved abundance can be considered an average of abundances probed at different pressure levels in the photosphere. Therefore, we expect to find a median retrieved abundance close to the expected value in the middle of the photosphere. More generally, we would at least expect the retrieved abundance to be within the range of top-of-atmosphere and bottom-of-atmosphere abundances set by the thermo-chemical equilibrium grids. Here, we can see photospheric agreement around 10 bar for CO and 3 bar for Na + K. A similar result is shown for NH$_3$ where the median retrieved abundance is in range of its equilibrium abundance in the photosphere. 

There is a distinction for our retrieved H$_2$O and CH$_4$ which show a slightly depleted abundance as compared to their equilibrium predictions. H$_2$O, CH$_4$ and CO are plotted together in Figure \ref{fig:profile and contribution}c as their chemical mixing influences the photospheric abundances we observe. We know from \citet{Oppenheimer1998} and \citet{Noll1997} that there is unexpectedly high amounts of CO in the atmosphere of Gl 229B based on spectral features, particularly at 4.7 $\mu$m, which suggest disequilibrium chemistry. A slightly depleted abundance of H$_2$O and CH$_4$ could be a result of strong vertical mixing that also causes CO abundance in excess of thermo-chemical equilibrium predictions in the photosphere. This is due to mixing timescales being shorter than the CO $\rightarrow$ CH$_4$ chemical timescale \citep{Fegley1996, Marley2015}, where the net reaction for this conversion in Gl 229B is 
\begin{equation}
CO + 3H_2 \rightarrow CH_4 + H_2O
\end{equation}
\citep{Visscher2011}. We discuss further causes for depleted oxygen abundance in Section \ref{sec:chemistry}. In general, since we expect chemical disequilibrium in the atmosphere of Gl 229B and we are freely retrieving gas abundances (as opposed to retrieving thermo-chemical equilibrium abundances), it is not surprising that our abundances for H$_2$O and CH$_4$ differ slightly from these chemical grids.

\begin{figure*}[ht!]
\centering
\plotone{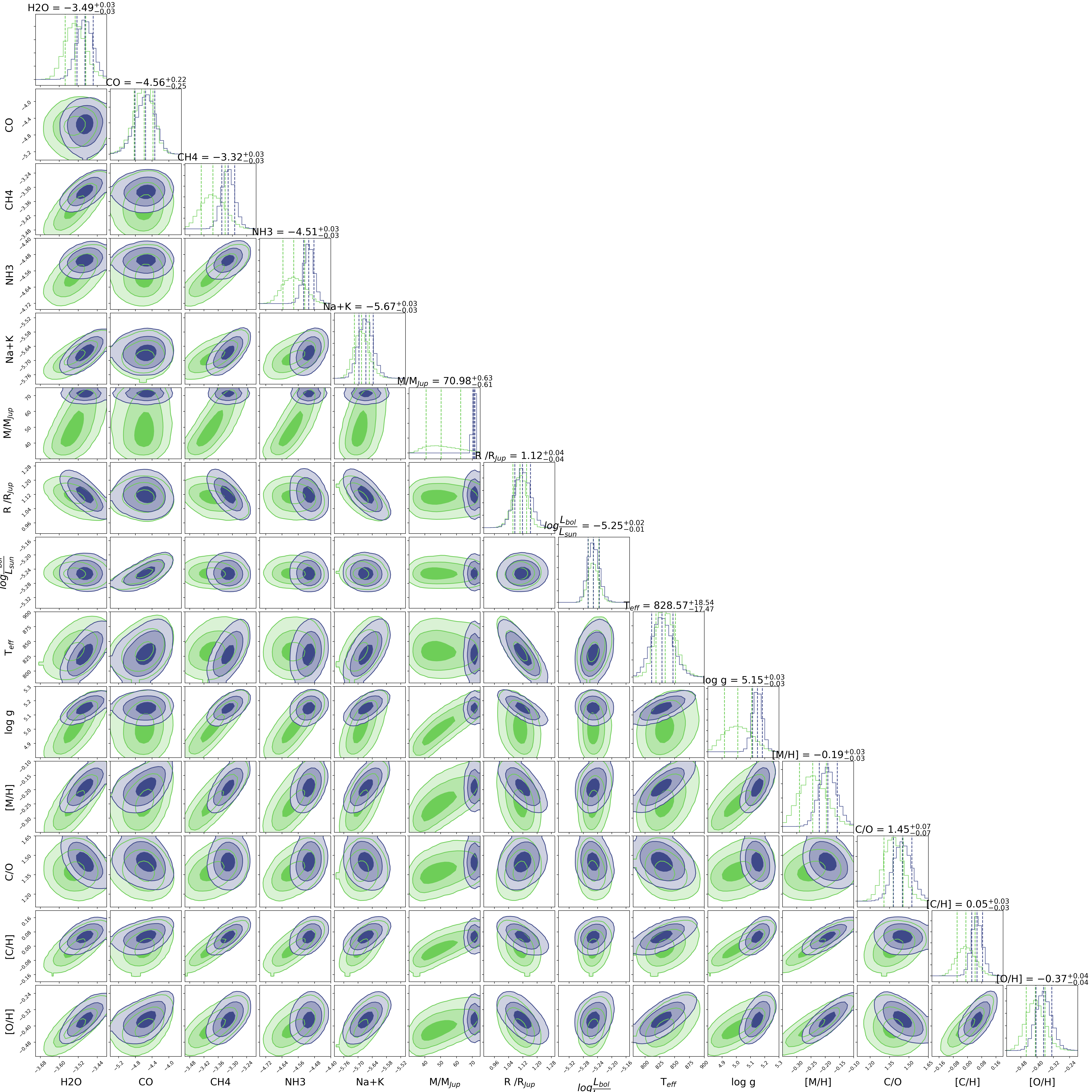}
\caption{\textit{Best Fit Model with a Mass Constraint} Posterior probability distributions for retrieved gas abundances and derived fundamental parameters of the mass constrained model shown in blue as compared to probability distributions from the best fit model (green). As before, marginalized posteriors are plotted on the main diagonal with 2D correlations between each parameters shown. Gas abundances are given in units of dex.
\label{fig:mass prior corner}}
\end{figure*}

\begin{figure*}[ht]
\centering
\plotone{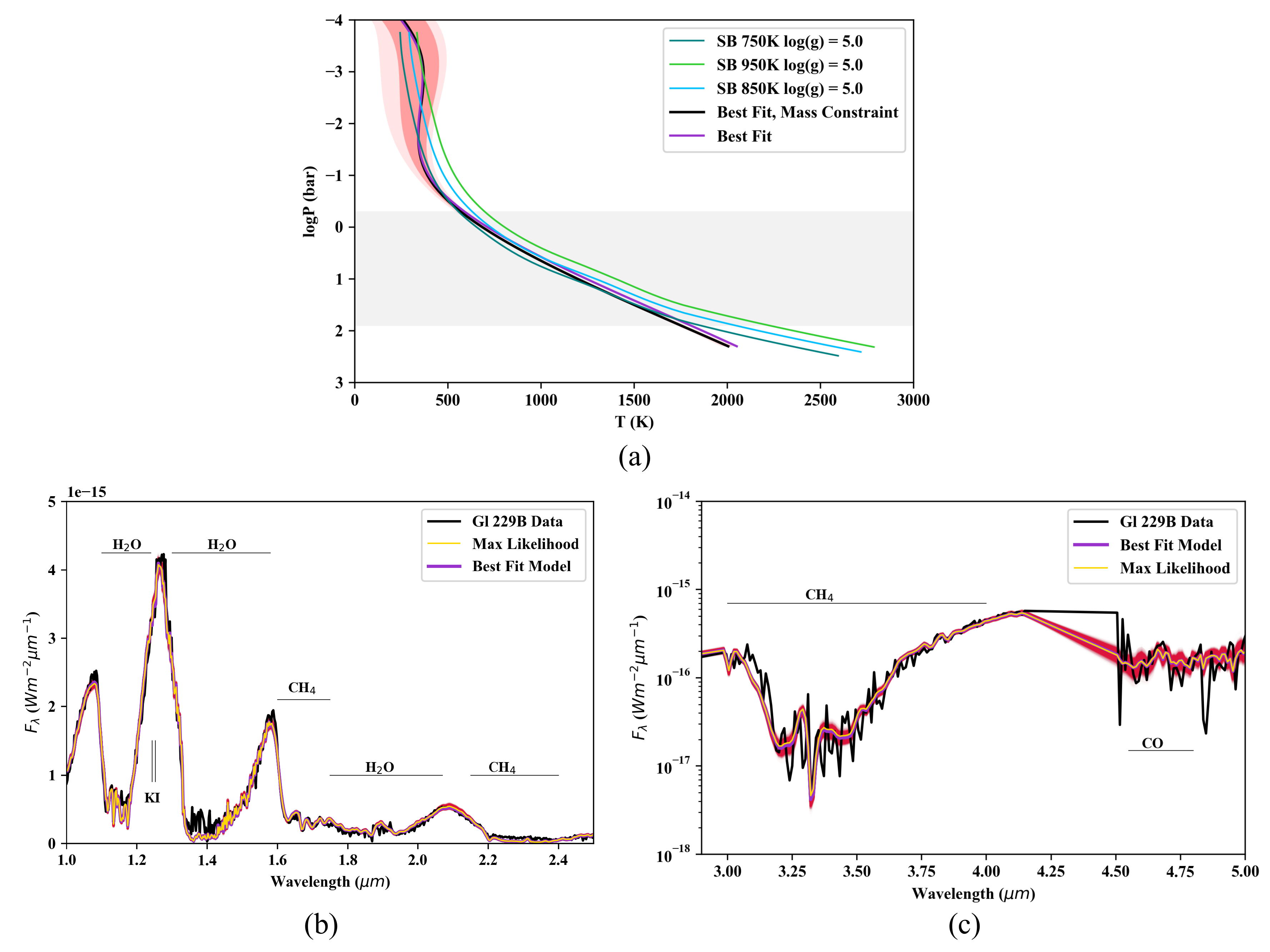}
\caption{\textit{Best Fit Model with a Mass Constraint} (a) Retrieved temperature-pressure profile for the mass constrained cloudless model with Allard alkalies as compared against Sonora model profiles (blue and green) and the retrieved profile from our best fit model (purple). We, again, show the approximate location of the photosphere as determined by our contribution function with the inclusion of a light grey panel. Figures (b) and (c) show the maximum likelihood retrieved spectrum for the mass constrained model in yellow compared against observed data (black) and the maximum likelihood retrieved spectrum from the best fit model (purple). In red, we show flux uncertainty on this retrieved spectrum.
\label{fig:mass prior spec}}
\end{figure*}

\subsubsection{Retrieved Spectrum vs. Forward Model Fit}
Figure \ref{fig:retrieved spectra}(a,b) show our retrieved spectrum as compared to the observed data. The retrieved spectrum fits the near-infrared portion of the observed spectrum well even though it struggles to reach the top of the flux peaks in the J and H spectral bands. The retrieved spectrum is able to fit the methane features at 1.63, 1.67 and 1.71 $\mu$m particularly well. It also fits the narrow absorption features due to water on the shortward side of the H-band peak. There is also a good fit to the top of the flux peak in K-band. The choice of alkalies had a significant impact on how our model fit the K-I doublet in the J spectral band. In particular, we find the use of Allard alkalies fits this K-I doublet much better than a cloudless model with Burrows alkalies. This difference in the retrieved spectrum J-band between the Allard and Burrows cloud models is likely a result of how pressure broadening for these lines is calculated for objects in this temperature regime. The retrieved spectrum does a good job of fitting the fundamental methane absorption feature of the L spectral band, particularly in the region from 3.6-4.1 $\mu$m. The retrieved spectrum also fits the general shape of the M spectral band data and attempts to fit the CO feature at 4.67 $\mu$m, providing a generally good fit within uncertainty bounds. This could be due to difficulty fitting disequilibrium abundances of CO, as CO is also the most poorly constrained of all the gases in our model. It is important to note that the narrow feature around $\sim$4.8 $\mu$m is due to an incorrect removal of a telluric line \citep{Noll1997} and therefore we do not expect our model to fit this.

Figure \ref{fig:retrieved spectra}(c,d) shows our retrieved spectrum and observed data in comparison to Sonora model synthetic spectra that bracket our retrieved $T_{\rm eff}$ . We find that the observed spectrum is best fit by the 850K Solar metallicity model, overall, but note that it does not reach the flux peak in the L-band. We also point out that none of the Sonora models seem to adequately fit the observed M-band data. While model spectra can provide good spectral fits in certain bands, no single model can simultaneously fit J, H, K, L and M spectral bands as well as our best fit retrieval model spectrum.

\begin{deluxetable*}{lcccccccc}[ht]
\tablenum{6}
\tablecaption{Comparison of Fundamental Parameters for Gliese 229B}
\tabletypesize{\footnotesize}
\tablehead{\colhead{Source} & \colhead{Radius ($R_{\rm Jup}$ )} & \colhead{Mass ($M_{\rm Jup}$ )} & \colhead{log g} & \colhead{$T_{\rm eff}$  (K)} & \colhead{log($L_{\rm Bol}$ /L$_{\rm Sun}$)} & \colhead{C/O} & \colhead{$[M/H]$} & \colhead{Age (Gyr)}}
    \startdata
    This Paper\tablenotemark{a} & 0.94 $\pm$ 0.15 & 41 $\pm$ 24 & 4.96 $\pm$ 0.46 & 927 $\pm$ 79 & -5.21 $\pm$ 0.05 & ... & 0.0 & 0.5-10 \\
    This Paper\tablenotemark{b} & $1.10 \pm 0.04$ & $50^{+12}_{-10}$ & $5.01^{+0.10}_{-0.09}$ & $834^{+17}_{-15}$ & $-5.25^{+0.02}_{-0.01}$ & $1.38^{+0.08}_{-0.07}$ & $-0.24 \pm 0.05$ & ... \\
    This Paper\tablenotemark{c} & $1.12 \pm 0.04$ & $71 \pm 1$ & $5.15 \pm 0.03$ & $829^{+19}_{-17}$ & $-5.25^{+0.02}_{-0.01}$ & 1.45 $\pm$ 0.07 & $-0.19 \pm 0.03$ & ... \\ 
    Naka95 & ... & 20-50 & ... & $<$1200 & -5.398 & ... & ... & 0.5-5 \\
    Naka15 & ... & 30-38 & 4.87 $\pm$ 0.12 & 825 $\pm$ 25 & ... & ... & 0.13 $\pm$ 0.07 & 1-2.5 \\
    Legg02 & ... &  $>$7 & 3.5 $\pm$ 0.5 & 1000 $\pm$ 100 & -5.21 $\pm$ 0.02 & ... & -0.5 & 0.016-0.045 \\
    Legg99 & ... & 25-35 & ... & 900 & -5.18 $\pm$ 0.04 & ... & ... & 0.5-1 \\
    Alla96 & ... & 40-58 & 5.3 $\pm$ 0.2 & 1000 & -5.21 $\pm$ 0.1 & ... & ... & 0.5-5 \\
    Matt96 & ... & ... & ... & 913 & -5.194 & ... & ... & ... \\
    Saum00 & ... & 15-73 & 5.0 $\pm$ 0.5 & 950 $\pm$ 80 & -5.21 $\pm$ 0.04 & ... & -0.3 $\pm$ 0.2 & $>$ 0.2 \\
    Bran20 & ... & 70 $\pm$ 5\tablenotemark{d} & 5.433 $\pm$ 0.033 & 1025 $\pm$ 15 & -5.208 $\pm$ 0.007\tablenotemark{e} & ... & ... & 7-10 \\
    Howe22 & 1.27 $\pm$ 0.03 & 41.6 $\pm$ 3.3 & $4.93^{+0.02}_{-0.03}$ & $869^{+5}_{-7}$ & ... & 1.13 $\pm$ 0.03 & -0.07 $\pm$ 0.03 & ...\\
\enddata
\tablenotetext{\tiny a}{SED}\vspace{-2ex}
\tablenotetext{\tiny b}{Best Fit Model}\vspace{-2ex}
\tablenotetext{\tiny c}{Mass Constrained Best Fit Model}\vspace{-2ex}
\tablenotetext{\tiny d}{This dynamical mass has since been updated in \citet{Brandt2021} with improved astrometry from Gaia EDR3 \citep{Gaia2021} to be 71.4 $\pm$ 0.6.}\vspace{-2ex}
\tablenotetext{\tiny e}{This value is taken from \citet{Filippazzo2015} which we have updated in this work.}\vspace{-2ex}
\tablerefs{Naka95: \citet{Nakajima1995}, Naka15: \citet{Nakajima2015}, Legg02: \citet{Leggett2002a}, Legg99: \citet{Leggett1999}, Alla96: \citet{Allard1996}, Matt96: \citet{Matthews1996}, Saum00: \citet{Saumon2000}, Bran20: \citet{Brandt2020}, Howe22: \citet{Howe2022}}
\label{tab:Gl229B Literature}
\end{deluxetable*}

\subsection{Best Fit Model with a Mass Constraint}\label{sec:best fit mass}
In this section, we present results from constraining the posterior probability range for the mass from 10 - 80 $M_{\rm Jup}$  to 70 - 72 $M_{\rm Jup}$  on our best fit model. The motivation for placing a prior on this model is the dynamical mass reported in \citet{Brandt2021} that predicts a companion mass for Gl 229A of 71.4 $\pm$ 0.6 $M_{\rm Jup}$ . Placing a prior on the model inherently increases model confidence so we do not use our Bayesian evidence parameter to compare with our best fit model (i.e. retrieval model without prior knowledge placed on retrieved parameters). However, we intend to use this model constraint to compare retrieved gas abundances and fundamental parameters to our best fit model in order to understand whether the photospheric chemistry can be explained by a 70 $M_{\rm Jup}$  object as well as a 50 $M_{\rm Jup}$  object in Section \ref{sec:mass}.

\subsubsection{Retrieved Gas Abundances and Fundamental Parameters}
Figure \ref{fig:mass prior corner} shows the posterior probability distributions for retrieved gas abundances and radius in addition to the derived log(g), $T_{\rm eff}$ , $L_{\rm Bol}$ , C/O ratio, [M/H], [C/H] and [O/H] for the mass constrained model. We list the values from Figure \ref{fig:mass prior corner} in Table \ref{tab:Retrieval results} for ease of reading.

The procedure to calculate $T_{\rm eff}$, log(g), C/O ratio and metallicity are described in Section \ref{sec:best fit}. Again, we have 1$\sigma$ agreement with the retrieved radius and derived log(g) to the values determined from the SED. The mass constrained model found a $T_{\rm eff}$  slightly cooler than the best fit model, however, we still have agreement within 1.2$\sigma$ to the SED value.

As compared to parameters of the best fit model, the $T_{\rm eff}$ , radius, and log($L_{\rm Bol}$ /$L_{\rm Sun}$) of the constrained model are all within 1$\sigma$ agreement. We have 1.5$\sigma$ agreement with log(g) values which is expected considering the higher mass of the constrained model. Additionally, we have retrieved abundances of H$_2$O, CO, Na and K in 1$\sigma$ agreement and CH$_4$ and NH$_3$ within 1.5$\sigma$ agreement between models. For an object of this temperature, we expect to see increased abundance in both CH$_4$ and NH$_3$ as a result of an increased log(g) as compared to CO, for example, which is thought to be less sensitive to gravity \citep{Zahnle2014}. Overall, we find the chemistry and temperature to be consistent between our best fit, cloudless model and a cloudless model with a constraint at the dynamical mass value.

\subsubsection{Temperature-Pressure Profile}
Figure \ref{fig:mass prior spec}(a) shows the retrieved T-P profile for the mass constrained cloudless model as compared to our best fit cloudless profile as well as Sonora grid models. As our derived $T_{\rm eff}$  is only slightly cooler ($\sim$ 5 K) for this mass constrained model than our best fit model, we find the Sonora Solar-metallicity log(g) = 5.0, 850 K model fits well throughout the top of the photosphere (1 - 5 bar). At pressures $>$ 10 bar and above the photosphere we have better agreement with the 750 K model. The 1$\sigma$ and 3$\sigma$ bounds, which were previously noted to be small for the best fit model, are shown to almost disappear below 1 bar. This is due to increased model confidence as a result of the prior constraint placed on mass. However, we still see good agreement to the log g = 5.0, 850 K model throughout the top of the photosphere until about P = 5 bar where the Sonora model diverges from our profile to slightly warmer temperatures. Above the photosphere, we have agreement within 3$\sigma$ to both 750 K and 850 K models. In contrast, the mass constrained model profile is approximately 100 K cooler throughout the photosphere than the Solar-metallicity log g = 5.0, 950 K model.

As shown in Figure \ref{fig:mass prior spec}(a), the best fit profile is nearly identical to the mass constrained profile, only diverging to slightly warmer temperatures ($\sim$ 10 K) toward the bottom of the photosphere (around 5 bar).

\begin{deluxetable*}{lcccccccc}
\tablenum{7}
\tablecaption{Comparison of Fundamental Parameters for Gliese 229A}
\tablehead{\colhead{Source} & \colhead{Radius ($R_{\rm Sun}$)} & \colhead{Mass ($M_{\rm Sun}$)} & \colhead{log g} &  \colhead{$T_{\rm eff}$ (K)} & \colhead{log($L_{\rm Bol}$ /$L_{\rm Sun}$)} & \colhead{C/O} & \colhead{$[M/H]$} & \colhead{Age (Gyr)}}
    \startdata
    Legg02 & ... & 0.30-0.45 & 3.75 $\pm$ 0.25 & 3700 & -1.29 & ... & -0.6 $\pm$ 0.1 & 0.016-0.045 \\
    Moul78 & ... & ... & 4.75 & 3610 & ... & ... & 0.15 $\pm$ 0.15 & ... \\
    Tsuj14 & 0.496 & 0.524 & 4.77 & 3710 & ... & ... & ... & ...\\
    Naka15 & ... & ... & 4.77 & ... & ... & 0.68 $\pm$ 0.12 & 0.13 $\pm$ 0.07 & ... \\
    Schi97 & ... & ... & 4.7 & 3330 & ... & ... & -0.2 $\pm$ 0.4\tablenotemark{a} & ... \\
    Gaid14 & 0.53 $\pm$ 0.05 & 0.56 $\pm$ 0.07 & ... & 3800 $\pm$ 100 & -1.30 $\pm$ 0.1 & ... & 0.12 $\pm$ 0.10\tablenotemark{a} & ...\\
    Neve14 & ... & 0.58 $\pm$ 0.03 & ... & 3630 $\pm$ 110 & ... & ... & -0.03 $\pm$ 0.09\tablenotemark{a} & ... \\
    Bran21 & ... & 0.579 $\pm$ 0.007 & ... & ... & ... & ... & ... & 2.6 $\pm$ 0.5\tablenotemark{b}\\
\enddata
\tablenotetext{\tiny a}{Metallicity reported as [Fe/H].}\vspace{-2ex}
\tablenotetext{\tiny b}{This age value is calculated in \citet{Brandt2020} based on stellar activity but they adopt a prior age of the system 1-10 Gyr due to disagreement between the derived stellar and brown dwarf ages.}\vspace{-2ex}
\tablerefs{Legg02: \citet{Leggett2002a}, Moul78: \citet{Mould1978}, Tsuj14: \citet{Tsuji2014}, Naka15: \citet{Nakajima2015}, Schi97: \citet{Schiavon1997}, Gaid14: \citet{Gaidos2014}, Neve14: \citet{Neves2014}, Bran21: \citet{Brandt2021} }
\label{tab:Gl229A Literature}
\end{deluxetable*}

\subsubsection{Retrieved Spectrum vs. Best Fit Model}
The retrieved spectrum for the mass constrained model is shown in Figure \ref{fig:mass prior spec}(b, c). As compared to the spectral fit from our best fit model, a cloudless model with a mass prior around 70 $M_{\rm Jup}$ fits the observed spectrum for Gl 229B just as well. These nearly identical fits show very good agreement throughout the near infrared and into the L and M spectral bands with both model spectra being able to fit the notable water, methane and carbon monoxide features. We do note that our mass constrained retrieved spectrum shows a marginally worse fit to the top of the flux peak in the J and H bands than the best fit retrieved spectrum fit but maintains a similarly good fit to the top of the flux peak in the K-band. The only notable difference between these model fits is the K-I doublet in the J-band where the mass constrained model predicts a slightly shallower feature than exists in the observed data. Otherwise, we find a similarly good spectral fit overall.

\section{Fundamental Parameter Comparison of Gl 229B} \label{sec:discussion}
\subsection{The Mass of Gl 229B}\label{sec:mass}
Of all fundamental parameters to consider, constraining the mass of Gl 229B has been a top priority as an updated dynamical mass of 71.4 $\pm$ 0.6 $M_{\rm Jup}$ reported in \citet{Brandt2021} is at odds with several evolutionary model predictions that placed Gl 229B in the 30-55 $M_{\rm Jup}$ range \citep{Allard1996, Marley1996, Nakajima1995, Nakajima2015}. While it is difficult to determine a precise age for Gl 229A, gyrochronology, thin disk kinematics and coronal and chromospheric activity all disfavor an old age, placing the Gl 229 system at an intermediate age of 2 - 6 Gyr \citep{Brandt2020}. Evolutionary models \citep[ex:][]{Allard2001, Saumon2008, Phillips2020, Burrows1997} predict that a T dwarf with a mass of 71.4 $\pm$ 0.6 $M_{\rm Jup}$ would have to be 4 - 7 Gyr older than the predicted age of Gl 229A to have cooled to an observable log($L_{\rm Bol}$/$L_{\rm Sun}$) $\approx$ -5.2. As a result, best fit evolutionary models place Gl 229B at or below 55 $M_{\rm Jup}$ in order to be consistent with this age-mass-luminosity relation. While our best fit model independently deduced a mass of 50$^{+12}_{-9}$ $M_{\rm Jup}$, we also present results using our best fit model (cloudless, Allard alkalis) with a constraint on the posterior probability space for mass from 1 - 80 $M_{\rm Jup}$ to 70 - 72 $M_{\rm Jup}$ . 

From our results in Section \ref{sec:best fit mass} with the constrained mass prior, we see an increase in all molecular abundances as compared to our best fit model. However, the abundances of the mass constrained model are all within or close to the 1$\sigma$ bounds of the best fit model. We find no difference in any retrieved abundance worth mentioning. There is an increase in our calculated log(g) from 5.01$^{+0.10}_{-0.09}$ to 5.15 $\pm$ 0.03 which is expected due to the increased mass. However, we do see a relative increase in retrieved radius from 1.10 $\pm$ 0.04 $R_{\rm Jup}$ to 1.12 $\pm$ 0.04 $R_{\rm Jup}$ which is unexpected since brown dwarfs are predicted to contract with age. However, these retrieved radii agree within 1$\sigma$ so while this result is unexpected it is not unreasonable. Retrieved and calculated values for both models are listed in Table \ref{tab:Retrieval results}.

While our model places no constraint on the age of Gl 229B, we present results showing the plausibility that the spectrum and atmospheric chemistry can be fit consistently well by a 70 $M_{\rm Jup}$ model as it can by a 50 $M_{\rm Jup}$ model.

\subsection{$L_{\rm Bol}$, Radius, {$T_{\rm eff}$}  and log(g) of Gl 229B}
In Table \ref{tab:Gl229B Literature} we list our SED derived and retrieved fundamental parameters as compared to values from the literature for Gl 229B. We find that our semi-empirical SED derived $L_{\rm Bol}$, radius, mass and log(g) agree within 1$\sigma$ to our best fit model retrieved parameters while $T_{\rm eff}$ agrees within 1.2$\sigma$.

Comparing our retrieval-derived log(g) to values published in the literature, we find agreement within 1$\sigma$ between our best fit model and all literature predictions except \citet{Leggett2002a} and \citet{Brandt2020}, however, we note that the published log(g) from \citet{Leggett2002a} is in 2$\sigma$ disagreement with all other published values due to the young age estimate of the system. Our mass constrained model log(g) is in agreement within 1$\sigma$ with \citet{Allard1996} and \citet{Saumon2000} and within 2.5$\sigma$ to \citet{Nakajima2015}. Our constrained model is not in agreement with the reported log(g) from \citet{Brandt2020}, however, the reported uncertainties for both values are notably small.

 The $T_{\rm eff}$ we derive for both the best fit and mass constrained model is cooler than all literature values except \citet{Nakajima2015}, which agrees within 1$\sigma$. We retrieved the same $L_{\rm Bol}$ for both constrained and best fit models and find its value is slightly smaller than those reported in the literature but still see 2$\sigma$ agreement to nearly all models, despite small uncertainties.

\begin{figure*}[ht!]
\hspace*{-0.5cm}
\includegraphics[scale=0.27]{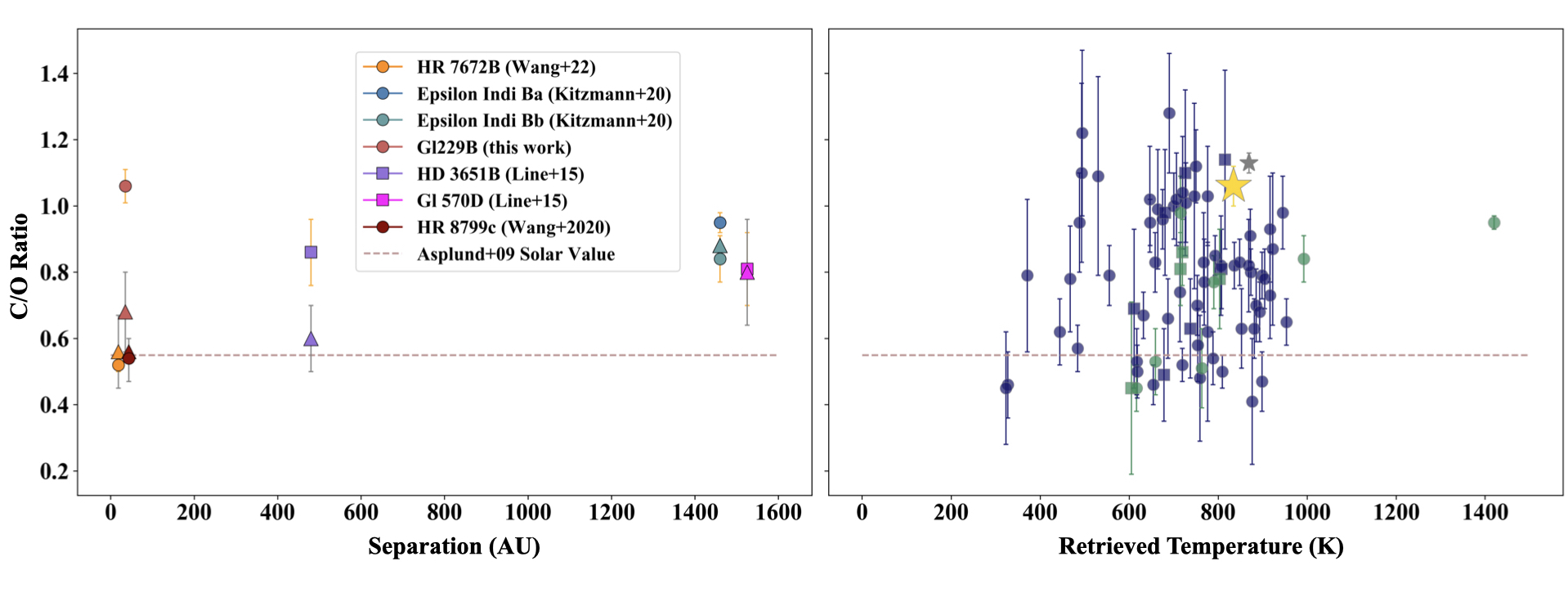}
\vspace{-2ex}
\caption{In the figure on the left, we show retrieved C/O values for objects with a main sequence companion with known C/O. Substellar objects are marked by a circle or square while their stellar primaries are marked by a triangle. On the right, we show a sample of brown dwarfs with retrieved C/O taken from \citet{Line2015, Line2017, Gonzales2020, Zalesky2019, Zalesky2022, Kitzmann2020, Howe2022}. Points in dark blue are isolated objects while points shown in green are companions. Gl 229B results from this work are marked by a yellow star whereas Gl 229B results from \citet{Howe2022} are marked by a grey star. Across both plots, objects marked by a circle denote reported bulk C/O ratio while squares denote C/O reported as CH$_4$/H$_2$O. The dashed line in both plots shows Solar C/O ratio of 0.55 taken from \citet{Asplund2009}.
\label{fig:all retrieved T dwarfs}}
\end{figure*}

\section{Gl 229B Compared to Other Retrieved T Dwarfs}\label{sec:t comp}
In this section, we place our best fit retrieval of Gl 229B in context with previous retrieval work on T dwarfs. Initially, \citet{Line2015} presented retrieval results of benchmark T dwarfs Gl 570D and HD 3651B which was later expanded upon in \citet{Line2017} with retrievals of a total of eleven T dwarfs. \citet{Zalesky2019} focused their work on ultra-cool dwarfs with a sample of 14 objects, six of which were late-T type dwarfs, and expanded on this work in \citet{Zalesky2022} with a sample of 50 T7-T9 type dwarfs. Additionally, \citet{Kitzmann2020} presented a retrieval study on the $\epsilon$ Indi Bab system, a system with two T-type companions. Finally, we examine our work in context with the retrieval of Gl 229B done by \citet{Howe2022}.

Consistent with \citet{Line2015, Line2017}, we do not include CO$_2$ or H$_2$S in our final model since preliminary results showed their abundances to be unconstrained. However, with the use of L and M spectral band data, we are able to put upper and lower bound constraints on CO abundance where there previously were none. From the results listed in Table \ref{tab:Retrieval models}, we also find that the data does not justify the addition of an optically thick cloud. We extend this result from \citet{Line2015, Line2017} to include optically thin clouds and further confirm that the data does not support the inclusion of condensates in the observable atmosphere as expected for this spectral type \citep[e.g.][]{Gao2020}.

From the sample in \citet{Line2017}, we focus our retrieved parameter comparison to objects within 100 K of our retrieved $T_{\rm eff}$  = 834$^{+17}_{-15}$ K for Gl 229B: 2MASS J$00501994-3322402$ ($T_{\rm eff}$  = 815$^{+20}_{-27}$ K), 2MASSI J$0727182+171001$ ($T_{\rm eff}$  = 807$^{+17}_{-19}$ K), 2MASSI J$1553022+153236$B ($T_{\rm eff}$  = 803$^{+16}_{-27}$ K) and 2MASS J$07290002-3954043$ ($T_{\rm eff}$  = 737$^{+21}_{-25}$ K). Our retrieved log(g) and radius are within 1$\sigma$ agreement to that of 2MASS J$00501994-3322402$ and 2MASSI J$0727182+171001$, only. We have 1$\sigma$ agreement on retrieved abundances of H$_2$O and CH$_4$ for 2MASS J$00501994-3322402$ and H$_2$O, CH$_4$ and NH$_3$ for 2MASSI J$0727182+171001$. We retrieve lower abundances of Na and K as compared to both objects which is likely due to the difference in alkali line opacities used in this work than \citet{Line2017}.

As the selected T dwarfs in \citet{Zalesky2019} are of type T8 or later, we have no direct comparison to focus on so we instead turn to \citet{Zalesky2022}. We discuss the objects in their sample close in temperature ($<$50 K difference) to that of our best fit for Gl 229B: WISE J$024124.73-365328.0$ ($T_{\rm eff}$ = 836 $\pm$ 4 K), WISE J$112438.12-042149.7$ ($T_{\rm eff}$ = 848$^{+26}_{-10}$ K) and WISEPC J$221354.69+091139.4$ ($T_{\rm eff}$ = 852$^{+12}_{-11}$ K). We have 1$\sigma$ agreement on retrieved log(g) and radius for WISE J$024124.73-365328.0$ and WISE J$112438.12-042149.7$ and 1.2$\sigma$ agreement with WISEPC J$221354.69+091139.4$. Since \citet{Zalesky2022} independently retrieves abundances for sodium and potassium, instead of tying their abundance ratio to solar value as we do here, we will only compare abundances for the constrained gases they report. While we do see slightly higher abundances of CH$_4$ and NH$_3$ in our model as compared to all three of these objects, we still have 1$\sigma$ agreement across H$_2$O, CH$_4$ and NH$_3$ abundance for WISE J$024124.73-365328.0$ and WISE J$112438.12-042149.7$. For J$221354.69+091139.4$, we have 1$\sigma$ agreement with H$_2$O and CH$_4$ abundance and 1.2$\sigma$ agreement with NH$_3$.

While the $\epsilon$ Indi system consists of an early T1.5 and T6 dwarf as a binary companion to a stellar type primary, we note that the T6 dwarf is too warm for a direct chemical comparison to Gl 229B. However, we are consistent with both \citet{Zalesky2019, Zalesky2022} and \citet{Kitzmann2020} in reporting cloud-free best fit models. We discuss these retrievals further in Section \ref{sec:comp chem} as they relate to developing C/O ratio and metallicity trends in retrieval work.

Finally, we place our Gl 229B retrieval in context with that done in \citet{Howe2022} which utilizes the same spectrum as we do in this work with the addition of 0.8 - 1.0 $\mu$m optical data from \citet{Schultz1998}. This work employs the retrieval code APOLLO (\citet{Howe2017}) and uses opacities from \citet{Freedman2014}. While \citet{Howe2022} reports uniformly small uncertainties on all retrieved gases and fundamental parameters, we do still have 1$\sigma$ agreement with our best fit model to retrieved abundances for CH$_4$ and NH$_3$ and retrieved log(g) and mass. We also have 2$\sigma$ agreement with retrieved temperature. However, we do see disagreement between retrieved abundances of H$_2$O, CO and Na + K where our model finds lower abundances for all three parameters on the order of 3$\sigma$ or greater. This could be due to differences in line lists used in these works as well as differences in underlying assumptions and biases between the two retrieval codes.

\section{C/O and Metallicity of Gl 229B}\label{sec:chemistry}
\subsection{C/O and Metallicity of Gl 229B as Compared to its Primary}
In Table \ref{tab:Gl229A Literature}, we list reported metallicities for Gl 229A which, similar to reported values for Gl 229B, range from subsolar to supersolar. For both best fit and mass constrained retrieval models, we find a subsolar metallicity which is within 1$\sigma$ and 2$\sigma$ of the value reported for Gl 229A in \citet{Schiavon1997} and \citet{Neves2014}, respectively. However, our derived metallicity skews subsolar due to retrieved oxygen abundances that are lower than expected for both models. Subsequently, we use carbon as a tracer for overall object metallicity (Gaarn et al. in prep) and calculate [C/H] = -0.01 $\pm$ 0.05 (best fit model) and [C/H] = 0.05 $\pm$ 0.03 (mass constrained model). A roughly Solar metallicity is in agreement with all published values for Gl 229A, except \citet{Leggett2002a} which reports a distinctly subsolar metallicity.

We list in Table \ref{tab:Gl229B Literature} the retrieval-based C/O ratio for Gl 229B and in Table \ref{tab:Gl229A Literature} the C/O ratio reported for Gl 229A from \citet{Nakajima2015}. \citet{Nakajima2015} determines a C/O ratio using carbon and oxygen abundances from \citet{Tsuji2014} and \citet{Tsuji2015}, a notoriously difficult task for M dwarfs due to molecular absorption features throughout their spectra. \citet{Tsuji2014} and \citet{Tsuji2015} use CO as an indicator of bulk carbon abundance and H$_2$O as an indicator of bulk oxygen abundance finding logA$_C$ = -3.27 $\pm$ 0.07 and logA$_O$ = -3.10 $\pm$ 0.02, respectively. The C/O ratio we find in this work is calculated based on the abundances of all retrieved carbon bearing molecules (CH$_4$, CO) as compared to all retrieved oxygen bearing molecules (H$_2$O, CO). We apply a correction to the oxygen abundance (30\% increase) to account for oxygen sequestered in silicate grains deeper in the atmosphere \citep{Line2015, Line2017, Zalesky2019, Zalesky2022}. For this correction, we assume the primary oxygen sink is enstatite (Mg$_2$Si$_2$O$_6$) and therefore account for the removal of three oxygen atoms for every magnesium silicate. We do this in an attempt to probe at a bulk, as opposed to atmospheric, C/O ratio for Gl 229B but note that this correction is still an approximation of the chemistry happening in the interior. The carbon and oxygen abundances of Gl 229A, B are listed in Table \ref{tab:C/O Ratio} for ease of comparison.

It is evident for both best fit and mass constrained models that our C/O ratios are not in agreement with that of the primary. Our best fit retrieved C/O is in agreement within 3$\sigma$ and our mass constrained, 4$\sigma$. Despite the median reported value for Gl 229A being supersolar, our C/O ratios are both $>$ 1. C/O ratio is theorized as tracer of formation mechanism such that stellar companions with elevated C/O ratios as compared to their primary likely formed due to core accretion in the disk \citep[e.g.][]{Madhusudhan2012, Oberg2011, Lodders2004, Madhusudhan2011, Konopacky2013}. While we do find higher C/O ratios in our models for Gl 229B, it is extremely unlikely that an object $\geq$ 50 $M_{\rm Jup}$ formed by core accretion in a disk around an M1V star (\citet{Schlaufman2018}, \citet{Bowler2015}, \citet{Mercer2020}). Furthermore, we compare the abundances of carbon and oxygen and find that the carbon values for both the best fit and mass constrained model agree with that of the primary within 1.2$\sigma$ and 1$\sigma$, respectively. We find this as evidence of coevality and consider the disagreement in oxygen abundance as a result of unexplained chemistry in the atmosphere of Gl 229B. As in \citet{Line2017}, we consider that a better understanding of the thermo-chemical mechanisms producing oxygen-bearing condensates is required to estimate a true oxygen abundance in cooler atmospheres. While magnesium silicates are the dominating oxygen sink in brown dwarf atmospheres, we hypothesize additional oxygen sinks such as iron-bearing species that could contribute to an oxygen-depleted atmosphere. However, an origin for such influential abundances of alternative metals is unknown.

To place our work in context with another retrieval analysis on Gl 229B, we turn, again, to the results from \citet{Howe2022}. As a result of increased abundance of oxygen-bearing absorbers as compared to our best fit model (see Section \ref{sec:t comp}), it is not surprising that they find a relatively smaller C/O ratio of 1.13 $\pm$ 0.03. It is also important to note that this work does not make an oxygen correction as we do, which would likely decrease this ratio even further. However, our work does align with \citet{Howe2022} in finding an overall supersolar C/O ratio, which, in their work, was speculated to be a result of unresolved disequilibrium chemistry in the atmosphere of Gl 229B.

\begin{deluxetable}{lccc}
\tablenum{8}
\tablecaption{Carbon and Oxygen Abundances}
\tablehead{\colhead{} & \colhead{logA$_C$} & \colhead{logA$_O$} & \colhead{C/O Ratio}}
    \startdata
    Gl 229A & $-3.27 \pm 0.07$\tablenotemark{a} & $-3.10 \pm 0.02$\tablenotemark{b} & 0.68 $\pm$ 0.12\tablenotemark{c}\\
    Gl 229B (BF) & $-3.35 \pm 0.05$ & $-3.49^{+0.05}_{-0.04}$ & $1.38^{+0.08}_{-0.07}$\\
    Gl 229B (BF)\tablenotemark{d} & $-3.35 \pm 0.05$ & $-3.38^{+0.05}_{-0.04}$ & $1.06^{+0.06}_{-0.05}$\\
    Gl 229B (MC) & $-3.29 \pm 0.03$ & $-3.45 \pm 0.04$ & $1.45 \pm 0.07$\\
    Gl 229B (MC)\tablenotemark{d} & $-3.29 \pm 0.03$ & $-3.34 \pm 0.04$ & $1.11^{+0.06}_{-0.05}$\\
    \enddata
    \tablecomments{(BF) indicates results from the best fit retrieval model while (MC) indicates results from the best fit model with a mass constraint.}
    \tablenotetext{a}{\citet{Tsuji2014}}\vspace{-2ex}
    \tablenotetext{b}{\citet{Tsuji2015}}\vspace{-2ex}
    \tablenotetext{c}{\citet{Nakajima2015}}\vspace{-2ex}
    \tablenotetext{d}{Corrected oxygen abundance and subsequent C/O ratio (see Section \ref{sec:chemistry}.}
\label{tab:C/O Ratio}
\end{deluxetable}

\subsection{C/O and Metallicity of Gl 229B Compared to Other Retrieved Isolated T Dwarfs}
There has been a noticeable trend developing toward supersolar C/O ratios in retrieval models of late T dwarfs to which our work on Gl 229B now contributes (Figure \ref{fig:all retrieved T dwarfs}). It is unclear whether this is a nod toward formation pathways, unresolved atmospheric chemistry in these types of objects or a bias in retrieval codes. This is certainly an active area of research that highlights the importance of brown dwarf companions, where studies can be anchored by the chemistry of the primary star (e.g. Gaarn et al. in prep). In this section, we return to the retrievals of \citet{Zalesky2019, Zalesky2022} and \citet{Line2017} to add our work to the growing chemical trends previously reported in these works.

Of the 11 T dwarfs retrieved in \citet{Line2017}, seven are isolated brown dwarfs with no potential for comparative chemistry to a Solar-type star. However, whether companion or not, a similar trend emerges for the entire sample of T dwarfs toward supersolar C/O ratio even after the implemented oxygen correction factor. Of these seven isolated objects, six were reported to have supersolar C/O ratios in range 0.63 - 1.14. It is important to make a distinction here that, as a result of unconstrained CO and CO$_2$, \citet{Line2017} calculated both C/O ratio and metallicity using CH$_4$ and H$_2$O abundances only. We note this because non-equilibrium chemistry, particularly with CO, could be a factor in the oxygen depletion seen in these retrievals. As a result, unconstrained abundances for CO and CO$_2$ could possibly be driving supersolar C/O ratios rather than unresolved chemistry or code biases.

Next, we consider our work against \citet{Zalesky2019, Zalesky2022} which initially presented retrieval models of a sample of 14 isolated late-type T dwarfs and Y dwarfs (T8-Y1), later expanded to a sample of 50 late-type T dwarfs (T7-T9). Of the six T dwarfs in the \citet{Zalesky2019} sample (T8-T9.5), they again find a trend toward supersolar C/O ratio in the late T regime with derived values in range 0.7 - 1.47 (with exception for one T9 object with a retrieved C/O of 0.57 $\pm$ 0.07). In this work, C/O ratio is calculated using H$_2$O, CH$_4$, CO and CO$_2$, despite only having upper limits on CO and CO$_2$. Consistent with \citet{Line2017}, \citet{Zalesky2019} also includes a C/O correction factor of 30$\%$. Unlike our work, \citet{Zalesky2019} reports slightly enhanced metallicities for this subset of T dwarfs. However, they report no apparent trend between high C/O and metallicity for late-T objects. Building on this initial study, \citet{Zalesky2022} also reports supersolar C/O ratios in range 0.65 - 1.28 for approximately 75$\%$ of their sample. For the comparative objects we discussed in \ref{sec:t comp}, $WISE J024124.73 - 365328.0$ has a retrieved C/O ratio of 0.82 $\pm$ 0.07, $WISE J112438.12 - 042149.7$ a value of $0.83^{+0.07}_{-0.08}$, and $WISEPC J221354.69 + 091139.4$, $0.63^{+0.12}_{-0.08}$. As our chemical abundances for Gl 229B were most similar to $WISE J024124.73 - 365328.0$ and $WISE J112438.12 - 042149.7$, it is not surprising that we find our C/O ratio closer in value to these objects than $WISEPC J221354.69 + 091139.4$. While our model for Gl 229B still reports a higher C/O ratio than all three of these temperature comparisons, we can still place our model alongside this apparent trend of supersolar C/O ratios in T dwarf retrievals. We visualize this trend in \ref{fig:all retrieved T dwarfs}.

\subsection{C/O and Metallicity of Gl 229B Compared to Other Companion Objects}\label{sec:comp chem}
The appeal of studying companion objects lies largely in the draw of understanding formation pathways. For brown dwarfs specifically, we hope that by comparing the atmospheric chemistry to that of its primary star (or brown dwarf companion) we can probe at the system age and make a determination on whether the two objects formed together and if they formed the same way (i.e. via gravitational fragmentation). As brown dwarf formation mechanisms are still unclear, studying co-moving systems provides one possible path to make advances on this topic.

In this section, we compare our findings on the Gl 229 system to the comparative chemistry of co-moving systems Gl 570 and HD 3651 \citep{Line2015, Line2017}, $\epsilon$ Indi \citep{Kitzmann2020}, HR 8799 \citep{Konopacky2013, Ruffio2021} and HR 7672 \citep{Wang2022}. We also plot C/O ratio of these brown dwarf companion systems in Figure \ref{fig:all retrieved T dwarfs}. These systems are of particular interest in relation to this work since they are all subjects of retrieval studies. Since Gl 570D, HD 3651B and $\epsilon$ Indi Bb are all late-T type objects, we would expect similar trends to emerge chemically (i.e. a high C/O ratio due to sequestered oxygen). However, we use the HR 8799 and HR 7672 systems as potential contrast as we wouldn't necessarily expect the same thermodynamic behavior in objects significantly hotter or cooler than Gl 229B. By placing the Gl 229 system in context with other retrieved co-moving systems, we can investigate trends in chemical abundance across companion objects with the ultimate goal of understanding underlying brown dwarf chemistry and formation.

\textit{Gl 570 (K4V+T7.5):} Gl 570D is an ideal companion candidate as it is widely separated from a well-studied main sequence star, Gl 570A. \citet{Line2015} reports metallicity and C/O ratio for the primary as $0.05 \pm 0.17$ and $0.81 \pm 0.16$, respectively. As previously discussed, \citet{Line2015, Line2017} applies a correction to the retrieved atmospheric oxygen abundance to probe at bulk abundance for Gl 570D, giving metallicity and C/O ratio values of $-0.15_{-0.09}^{+0.07}$ and $0.79^{+0.28}_{-0.23}$, respectively. While the spread for C/O spans solar to supersolar and metallicity subsolar to solar, there is agreement to 1$\sigma$ for these chemical parameters between Gl 570D and its primary which is in contrast to our model.

\textit{HD 3651 (K0V+T7.5):} HD 3651B is another widely separated companion to a well-studied main sequence star, HD 3651A. \citet{Line2015} reports metallicity and C/O ratio for the primary as $0.18 \pm 0.07$ and $0.62 \pm 0.11$, respectively. The oxygen corrected metallicity and C/O ratio of HD 3651B are $0.08_{-0.06}^{+0.05}$ and $0.89^{+0.21}_{-0.17}$, respectively. Clearly, this system shares a similarity to our work in that the primary star is of solar C/O and its brown dwarf companion is distinctly supersolar.

\textit{$\epsilon$ Indi (K4.5V+T1.5+T6.5):} The $\epsilon$ Indi system is another useful example where the main sequence primary can help put constraints on fundamental parameters. \citet{King2010} reports a metallicity measurement of $-0.2$ for the primary which is in agreement only with $\epsilon$ Indi Bb, the T6.5 dwarf, at a retrieved value of $-0.34^{+0.12}_{-0.11}$. The subsolar metallicity is predicted by the chemistry of the primary star, however, this work does not correct for condensation of oxygen-bearing condensates so we would expect better agreement after this assumption. A supersolar C/O ratio reported for both $\epsilon$ Indi Bb at $0.84 \pm 0.07$ and $\epsilon$ Indi Ba at $0.95 \pm 0.03$ is also predicted to be closer to solar after a correction. Similar to \citet{Line2015, Line2017} this work was not able to place constraints on CO, CO$_2$, or H$_2$S but does, in fact, calculate a C/O with all oxygen- and carbon-bearing molecules. Another interesting feature of this work, similar to ours, is that dynamical measurements predicted both brown dwarf companions to be $\sim$ 50-70 $M_{\rm Jup}$  \citep{Dieterich2018, Chen2022} while retrieval models place Ba and Bb at $50^{+7.8}_{-6.6}$ and $15^{+6}_{-4.6}$ $M_{\rm Jup}$ , respectively.

\textit{HD 7672 (G0+L4):} HD 7672 is another system in which the solar-type primary is well studied with reported abundances [Fe/H] = $-0.04 \pm 0.07$, [C/H] = $-0.08 \pm 0.05$, [O/H] = $0.15 \pm 0.06$ and C/O = $0.56 \pm 0.11$ \citep{Wang2022}. Retrieval derived abundance for the brown dwarf companion are reported as [C/H] = $-0.24 \pm 0.05$, [O/H] = $-0.19 \pm 0.06$ and C/O = $0.52 \pm 0.02$. This is an interesting divergence from our work where the C/O ratio is within 1$\sigma$ agreement to its primary but [C/H], which we use as a proxy for overall object metallicity, is subsolar. Since this brown dwarf companion is an L-type, we should not need to correct for oxygen-bearing condensates although \citet{Wang2022} does state that retrieved carbon and oxygen abundances are atmospheric (as opposed to bulk) and suggests that inefficient mixing could be a cause of these metallicity differences.

\textit{HR 8799 (A5/F0+bcde):} This system provides an interesting contrast to ours as this young star has four substellar companions with masses in range 5 - 13 $M_{\rm Jup}$  orbiting within 15 - 70 AU which is beyond the H$_2$O ice line in the disk \citep{Konopacky2013}. Initial studies from \citet{Konopacky2013}, found enhanced C/O ratio for the b, c companions as compared to the primary which is approximately solar at $0.54^{+0.12}_{-0.09}$, suggesting a core accretion scenario. However, \citet{Ruffio2021} concluded that all four planets have stellar C/O ratios, which could imply either core accretion or gravitational instability as a formation mechanism. This result is in agreement with retrieval results on HR 8799c, e \citep{Wang2020, Moll2020} but in disagreement with the retrieval study of \citet{Lavie2017} which found all four companions to be oxygen-enriched relative to their star. This opens up a particularly interesting discussion on the usefulness of C/O ratio as a formation diagnostic for objects near or above the deuterium burning limit. We include HR 8799c in Figure \ref{fig:all retrieved T dwarfs} as a contrast to higher mass, brown dwarf companions that show distinct supersolar C/O trends, as well as chemical non-uniformity to their primary star.

While C/O ratio has the potential to be a powerful formation diagnostic, the complications that arise in our attempt to determine bulk C/O ratios in substellar objects currently prevent us from using it as conclusive evidence. While we expect oxygen depletion in T-type objects, we still see unexpectedly incongruous C/O ratios for the Gl 229, HD 3651 and $\epsilon$ Indi systems. However, retrieval studies on Gl 570 and HD 7672 were able to match C/O ratio across companions within 1$\sigma$. Additionally, in the HR 8799 system where we initially predicted superstellar C/O ratios, retrieval results found substellar or stellar ratios. From these varied retrieval results so far, we do not see a unique chemical trend emerging (i.e. anomalously high C/O ratios for all T-type objects indicating unresolved chemistry in that temperature regime or anomalously high C/O ratios for all brown dwarfs suggesting a retrieval code bias). However, we also consider the possibility that both unresolved chemistry and code bias are contributing to the retrieval results we present.

\section{Conclusions} \label{sec:conclusion}
In this work we present an updated distance-calibrated SED and retrieval model of Gl 229B. Gl 229B is best fit by a cloudless model which is in agreement with previous retrieval work on T dwarfs \citep{Line2015, Line2017, Gonzales2020, Zalesky2022} as well as prior forward model predictions \citep[i.e.][]{Saumon2000, Leggett2002a, Allard1996}. We find that our retrieved mass and log(g) are consistent with evolutionary model predictions to within 1$\sigma$. However, we do find that our $T_{\rm eff}$  is slightly cooler than its SED-derived value which is a result of the retrieved radius being larger than predicted. We also find the use of Allard alkalies to provide a better spectral fit than Burrows, consistent with \citet{Gonzales2020}.

Additionally, we find that a cloud-free model with a prior constraint on mass near its dynamical value from \citet{Brandt2021} is capable of reproducing the chemistry and fit to the observed spectrum consistantly well as our best fit cloud-free model. Although we make no age determination, we find it plausible that Gl 229B is a 70 $M_{\rm Jup}$ object.

We discuss the implications of an anomalously high C/O ratio, especially as compared to its stellar value. We make a 30$\%$ correction for oxygen sequestered into condensates deep below the photosphere and still find a C/O ratio $>$ 1 for our best fit model. We find this work contributes to a trend in brown dwarf retrievals toward supersolar C/O ratios, particularly for T dwarfs. We also note that this high C/O ratio is due solely to “missing" oxygen in the atmosphere as we can match the abundance of carbon to its reported stellar value. As a result, we use [C/H] as a marker for overall metallicity and find that Gl 229B is of solar metallicity as the literature predicts. While we cannot unify the dissimilarity in C/O between Gl 229B and its primary, we find the agreement in carbon abundance a strong nod toward co-evalty. Overall, these results lead us toward more questions on the nature of brown dwarf atmospheric chemistry, formation, evolution and even biases in modelling approaches that we plan to explore in future work.

\section*{acknowledgments}
The data used in this publication were collected through the MENDEL high performance computing (HPC) cluster at the American Museum of Natural History. This HPC cluster was developed with National Science Foundation (NSF) Campus Cyberinfrastructure support through Award 1925590. This work was also supported by the National Science Foundation via awards 1909776 and AST-1909837.

\software{astropy \citep{astropy:2013}, \textit{Brewster} \citep{Burningham2017}, Corner \citep{corner}, \texttt{EMCEE} \citep{Foreman2013}, \texttt{PyMultinest} \citep{Buchner2014}, SEDkit \citep{Filippazzo2015}}

\bibliography{refs}
\bibliographystyle{aasjournal}

\appendix
\restartappendixnumbering
\section{Comments on Alkalis Used in Retrieval}\label{sec:appendix}
\begin{figure*}
\centering
\plotone{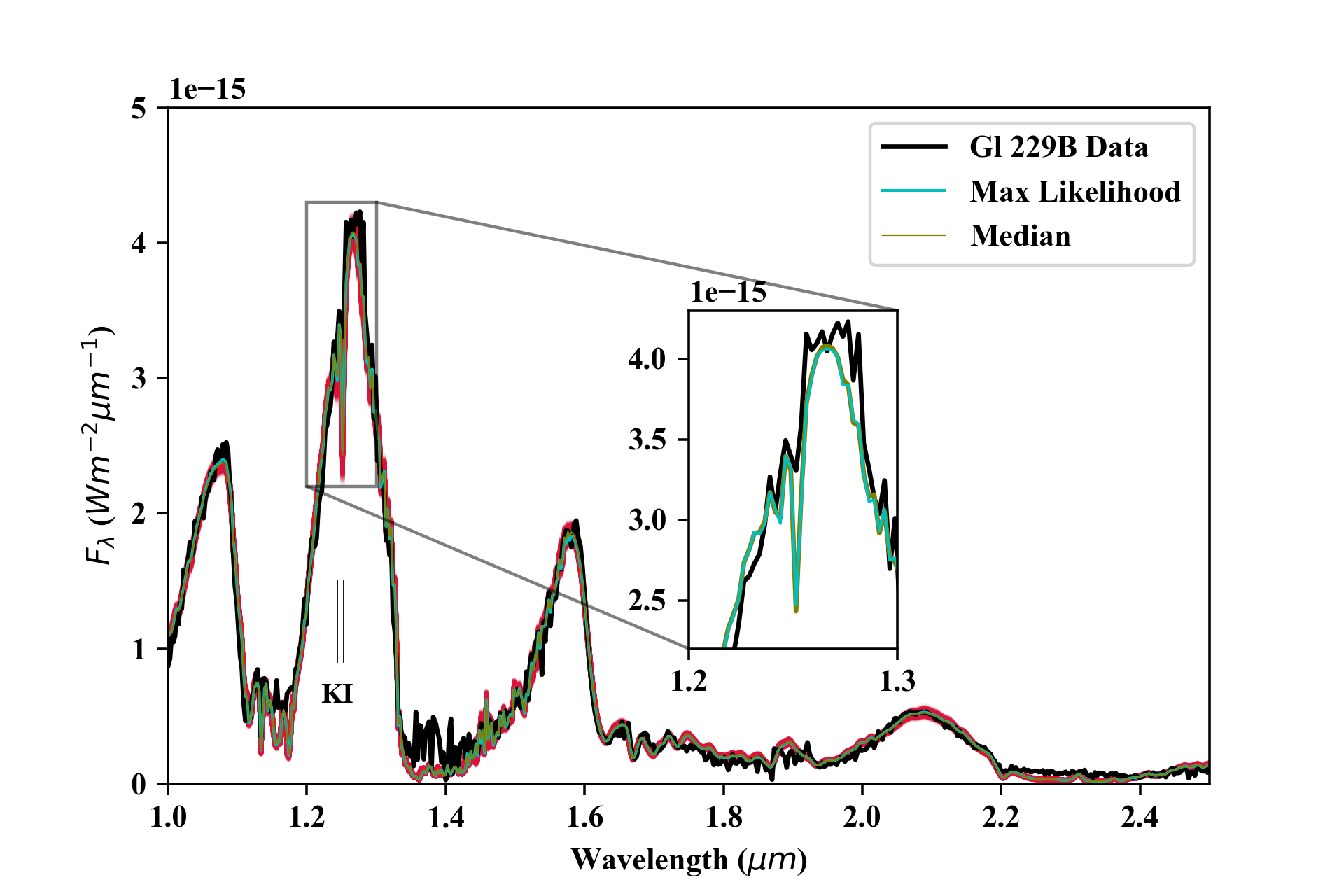}
\caption{This figure shows the results of a cloudless model with Burrows alkalies where the spectral fit to the K-I doublet in the J spectral band (around 1.25 $\mu$m) is severely overestimated. Observed data is shown in black whereas the maximum likelihood retrieved spectrum is in blue. This model used wavelength coverage 1.0 - 5.0 $\mu$m, however, we shorten the wavelength coverage in this figure to give a clearer view of this alkali feature. This serves as a comparison to our best fit model which uses Allard alkalies and shows an objectively good fit in this region.
\label{fig:burrows}}
\end{figure*}

The use of Allard opacities for T dwarf retrievals follows the work of \citet{Gonzales2020} that found that Burrows opacities produced retrieved alkali abundances that were incongruous with expected values due to rainout in T dwarfs. We also find the use of Allard opacities is strongly preferred by our model as determined by the Bayesian selection criterion where a cloudless model with Burrows opacities is strongly rejected as compared to our best fit model (Table \ref{tab:Retrieval models}). We confirm these results by the spectral fit of a cloudless model with Burrows opacities Figure \ref{fig:burrows}. It is clear that there is an extremely poor fit of the K-I feature in the J-band where the model predicted a much stronger absorption line than exists in the observed data. As discussed in Section \ref{sec:best fit}, we find that a cloudless model using Allard opacities can, in fact, fit this alkali feature.

\end{document}